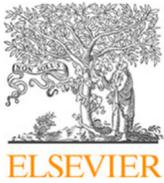
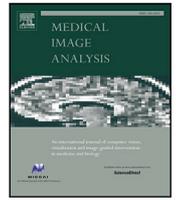

# Uncertainty quantification for White Matter Hyperintensity segmentation detects silent failures and improves automated Fazekas quantification

Ben Philps [a], Maria del C. Valdés Hernández [b],*, Chen Qin [c], Una Clancy [b], Eleni Sakka [b], Susana Muñoz Maniega [b], Mark E. Bastin [b], Angela C.C. Jochems [b], Joanna M. Wardlaw [b], Miguel O. Bernabeu [d], Alzheimer's Disease Neuroimaging Initiative (ADNI)[1]

[a] *School of Informatics, University of Edinburgh, Edinburgh EH8 9AB, United Kingdom*
[b] *Centre for Clinical Brain Sciences, Chancellor's Building, University of Edinburgh, Edinburgh EH16 4SB, United Kingdom*
[c] *Department of Electrical and Electronic Engineering & I-X, Imperial College London, London SW7 2AZ, United Kingdom*
[d] *Centre for Medical Informatics, Usher Institute, University of Edinburgh, Edinburgh EH16 4UX, United Kingdom*

## ABSTRACT

White Matter Hyperintensities (WMH) are key neuroradiological markers of small vessel disease present in brain MRI. Assessment of WMH is important in research and clinics. However, WMH are challenging to segment due to their high variability in shape, location, size, poorly defined borders, and similar intensity profile to other pathologies (e.g stroke lesions) and artefacts (e.g head motion). In this work, we assess the utility and semantic properties of the most effective techniques for uncertainty quantification (UQ) in segmentation for the WMH segmentation task across multiple test-time data distributions. We find UQ techniques reduce 'silent failure' by identifying in UQ maps small WMH clusters in the deep white matter that are unsegmented by the model. A combination of Stochastic Segmentation Networks with Deep Ensembles also yields the highest Dice and lowest Absolute Volume Difference % (AVD) score and can highlight areas where there is ambiguity between WMH and stroke lesions. We further demonstrate the downstream utility of UQ, proposing a novel method for classification of the clinical Fazekas score using spatial features extracted from voxelwise WMH probability and UQ maps. We show that incorporating WMH uncertainty information improves Fazekas classification performance and calibration. Our model with (UQ and spatial WMH features)/(spatial WMH features)/(WMH volume only) achieves a balanced accuracy score of 0.74/0.67/0.62, and root brier score (↓) of 0.65/0.72/0.74 in the Deep WMH and balanced accuracy of 0.74/0.73/0.71 and root brier score of 0.64/0.66/0.68 in the Periventricular region. We further demonstrate that stochastic UQ techniques with high sample diversity can improve the detection of poor quality segmentations.

## 1. Introduction

Cerebral small vessel disease (SVD) is a significant vascular contributor to stroke, cognitive decline, and dementia. One of its main neuroradiological signatures are White Matter Hyperintensities (WMH). They reflect brain tissue damage ranging from altered interstitial fluid mobility and water content to demyelination and axonal loss (Wardlaw et al., 2015), and appear as clusters with increased signal in the T2-weighted-based sequences from brain magnetic resonance imaging (MRI) scans, typically increasing in incidence with age. WMH foci are indicative of more widespread and subtle changes (Maillard et al., 2011; Valdés Hernández et al., 2014) and a predictor for increased stroke and mortality risks (Debette and Markus, 2010). Depending on brain location, WMH may have differing etiological substrates (Kim et al., 2008) and correlates (Wang et al., 2022; Jiménez-Balado et al., 2022; Gootjes et al., 2004; Griffanti et al., 2018). Hence, assessment of WMH burden, location, and correlation with clinical outcomes is important in clinical trials and epidemiological studies.

WMH burden in scans of elderly and SVD patients is typically visually assessed by neuroradiologists separately in the periventricular and deep white matter regions using Fazekas scores (Fazekas et al., 1987). However, despite the straightforwardness of this visual rating

---

\* Corresponding author.
*E-mail address:* M.Valdes-Hernan@ed.ac.uk (M.d.C. Valdés Hernández).
[1] This article uses data obtained from the Alzheimer's Disease Neuroimaging Initiative (ADNI) database (adni.loni.usc.edu). As such, the investigators within the ADNI contributed to the design and implementation of ADNI and/or provided data but did not participate in analysis or writing of this report. A complete listing of ADNI investigators can be found at: http://adni.loni.usc.edu/wp-content/uploads/how_to_apply/ADNI_Acknowledgement_List.jpg.

https://doi.org/10.1016/j.media.2025.103697
Received 22 November 2024; Received in revised form 6 May 2025; Accepted 13 June 2025
Available online 12 July 2025

scale, reliable assessment of WMH requires substantial expertise due to their heterogeneity in presentation and often simultaneous presence and sometimes coalescence of other pathologies of similar signal intensity characteristics (Ding et al., 2020). The gold standard for WMH computational assessment is manual delineation in fluid-attenuated inversion recovery (FLAIR) MRI. However, manual annotation of WMH is time-consuming, subject to inter-rater variability and requires substantial training time, as the boundaries between normal appearing white matter and the hyperintense tissue are generally ill-defined and difficult to delineate. Moreover, given the co-existence of WMH with features of similar appearance or even imaging artefacts, the identification of deep isolated WMH is sometimes subjective. These imply the existence of a distribution of multiple potential WMH segmentations and, by extension, Fazekas scores. Hence fast and reliable probabilistic automated techniques for WMH segmentation are essential.

Artificial Intelligence (AI) techniques have shown broad applicability to clinical (Monsour et al., 2022; Barnett et al., 2023) and research goals (Nenning and Langs, 2022) across a wide range of neuroimaging settings (Choi and Sunwoo, 2022) and disease types (Leming et al., 2023; Ruffle et al., 2023). In recent years a number of approaches have been developed to segment WMH lesions automatically with promising results (Balakrishnan et al., 2021). In particular, Convolutional Neural networks (CNNs) have yielded unprecedented performance (Litjens et al., 2017) in this segmentation task (Park et al., 2021; Mojiri Forooshani et al., 2022; Guerrero et al., 2018; Rachmadi et al., 2020; Zhao et al., 2021; Joo et al., 2022).

However, as most AI schemes, such techniques face reliability and robustness concerns that still hamper their trust and deployment (Cutillo et al., 2020; Zou et al., 2023a; Lim et al., 2019), particularly in the face of limited, biased or unrepresentative data sources. Large, population scale analysis of treatment effects and disease progression will likely increasingly rely on access to data collected across a multitude of scanning sites with accompanying variations in acquisition protocol (Borchert et al., 2023; Van Horn and Toga, 2014), or clinical imaging data (Ruffle et al., 2023; Gao et al., 2022) where image quality is typically lower. But while such data allows detecting complex patterns, this diversity in input covariates introduces clear robustness challenges for our models. Consequently, researchers and clinicians may wish to avoid delegating decision making to AI tools (Begoli et al., 2019), or to avoid AI generated information entirely when making decisions. Thus it is desirable to implement models that are not only robust but can quantify any uncertainties in the inferences made. Uncertainty quantification (UQ) is often touted as essential (Begoli et al., 2019) and a core tenement of improving the trust of such systems (Zou et al., 2023b). In the context of segmentation, this may be represented as a separate UQ output or a distribution of possible segmentations. This uncertainty should ideally reliably provide information on where a segmentation may be incorrect or the image is ambiguous (Czolbe et al., 2021), and estimate the true frequency of different spatially variant segmentations (Monteiro et al., 2020; Kohl et al., 2018).

UQ has been applied across a wide range of medical tasks (Yeung et al., 2023; Abdar et al., 2021; Oda et al., 2021; Tanno et al., 2021), imaging modalities (Lambert et al., 2024; Joy et al., 2021; Yeung et al., 2023) and data challenges from incomplete data to measurement variability (Seoni et al., 2023) and image denoising (Laves et al., 2020). UQ techniques typically modify our existing neural network models to produce a calibrated confidence evaluation along with the prediction of the segmentation model. They are commonly used for improving the quality of the predictions at inference time (You et al., 2023; Teja and Fleuret, 2021; Rahman et al., 2021; Zou et al., 2023b), such as uncertainty aware aggregation of predictions at different scales, improving segmentation when the target objects vary significantly in size and intensity (Oda et al., 2021), or to prevent overfitting to image noise in Bayesian deep image prior networks (Laves et al., 2020). For WMH segmentation, UQ could inform on ambiguous regions or model failures, capture unusual pathologies and generate multiple plausible segmentation possibilities. This improves the reliability of such models while effectively estimating the confidence of their outputs. Using uncertainty features for quality control has proved effective in various areas (Joy et al., 2021), with aleatoric uncertainty predicting segmentation error (Czolbe et al., 2021), or detecting poor quality predictions (Liang et al., 2022). More generally, as increasing interest in deploying neural networks in high risk scenarios increases, so does the importance of ensuring the trustworthiness of our models (Li et al., 2023b).

While many works have assessed UQ techniques, a comparison of the utility of different techniques for *downstream* clinical tasks is lacking (Czolbe et al., 2021; Zou et al., 2023a). In this work we compare multiple UQ techniques not only in terms of their visual utility in WMH assessment, but propose that UQ information can be used to enhance the prediction of the downstream clinical Fazekas score and for assessing the quality (e.g Dice score) of the WMH segmentations. We present 4 main contributions:

(1) We benchmark UQ methods for WMH segmentation and analyse the robustness and uncertainty map utility of each technique. We assess the Softmax entropy, MonteCarlo Dropout, Deep Ensembles, Evidential Deep learning, Explicit Heteroscedastic Classification (including Stochastic Segmentation Networks), and Probabilistic U-Net, demonstrating the added robustness from modelling both epistemic and aleatoric uncertainty, along with the increased detection of unsegmented small isolated WMH in uncertainty maps.

(2) We propose a novel method for Fazekas classification using spatial features extracted from WMH spatial probability and UQ maps, demonstrating that UQ map information can improve the prediction of the Fazekas score.

(3) We demonstrate that UQ features can aid in the identification of low quality segmentations.

(4) We conduct a qualitative analysis of the semantic information contained in UQ maps, examining images with small isolated clusters of WMH, stroke lesions and image artefacts.

## 2. Background

### 2.1. Uncertainty quantification overview

Uncertainty may be classified as epistemic (model uncertainty) or aleatoric (data uncertainty) (Abdar et al., 2021; Wimmer et al., 2023). Epistemic uncertainty reflects a lack of knowledge about the input, which is high for out-of-distribution (OOD) and unseen data and can thus be reduced with more data. For example, epistemic uncertainty may arise due to MRI acquisition characteristics changes or differences in cohort demographics. Aleatoric uncertainty, which is inherently irreducible, reflects the variability in the data (Abdar et al., 2021). For WMH this may include boundary delineation within the WMH "penumbra", possible misclassified foci, coexistence of WMH with other similar pathologies, confounding imaging artefacts and differences in the policy or definition of WMH used by annotators.

### 2.2. Uncertainty quantification techniques

This section outlines the task at hand and the principles behind UQ techniques benchmarked in this work. Given a dataset $D = \{x^{(n)}, y^{(n)}\}_{n=1}^{N}$ of brain MRI images $x$ and segmentation labels $y$, where $x$ is a pair of FLAIR and T1-weighted images and $y$ is a voxelwise binary segmentation mask, our task is to construct a model $f(x)$ with parameters $\theta$ that predicts $p(y|x, D; \theta)$. For training we consider $x$ and $y$ as individual 2D axial slices of brain MRI images, while for evaluation we combine slice predictions for 3D analysis.

*2.2.1. Softmax entropy*

Softmax entropy, a simple and commonly employed technique utilises the entropy of the Softmax distribution $\mathcal{H}[p(y|x, \mathcal{D}; \theta)]$ without requiring architectural changes or extra compute cost. It may effectively represent aleatoric uncertainty for in distribution data (Mukhoti et al., 2021). However it is essential to employ a loss function that rewards a calibrated model (e.g by utilising a proper scoring loss function such as cross entropy).

*2.2.2. Monte-Carlo dropout*

Bayesian neural networks represent model weights as a probability distribution. However computing the posterior predictive distribution analytically is intractable for a typical segmentation network. Hence, multiple approaches for approximating the posterior probability distribution have been proposed (Blei et al., 2017; Blundell et al., 2015; Gal and Ghahramani, 2016; D'Angelo et al., 2021) which re-frame bayesian inference as an optimisation problem.

Monte-Carlo dropout (Gal and Ghahramani, 2016), a quasi variational inference technique, uses dropout (Srivastava et al., 2014) at test time to generate stochastic model passes and is heavily used in the UQ literature. MC-Dropout approximates the Bayesian posterior predictive as:

$$p(y|x, \mathcal{D}) \approx \frac{1}{S} \sum_{s=1}^{S} p\left(y|x, \theta^{(s)}\right), \tag{1}$$

where $\theta^{(s)} \sim p(\theta|\mathcal{D})$. Since this term is intractable, it is approximated as $\theta^{(s)} = \theta \odot z^{(s)}$, where $z^{(s)}$ is a randomly generated binary dropout mask. Monte-Carlo Dropout has been widely applied to medical imaging (Lambert et al., 2024; Hu et al., 2019; Hellström et al., 2023; Laves et al., 2020), and has been used in WMH segmentation to improve robustness to downsampling and gamma correction augmentations that aim to mimic clinical data perturbations (Mojiri Forooshani et al., 2022).

*2.2.3. Deep ensembles*

Deep Ensembles (Lakshminarayanan et al., 2017) are a simple and typically state-of-the-art technique for UQ (Abdar et al., 2021), involving training multiple models (elements). There are many ensembling strategies (Ganaie et al., 2022). Here we focus on a simple strategy where multiple models are trained with different random weight initialisations and different shuffling orders per epoch. This leverages the non-convexity of the loss surface and the stochasticity of optimisation (Abe et al., 2022) to create diverse models which collectively outperform individual elements (Lee et al., 2015). Ensembles can provide well calibrated predictions (Lakshminarayanan et al., 2017), improve robustness to covariate shirt (Ovadia et al., 2019; Mehrtens et al., 2022) and improve WMH segmentation performance (Guo et al., 2022; Molchanova et al., 2025). Each element can be treated as a sample from an aggregate Bayesian model, which approximates the Bayesian posterior predictive distribution as a uniform mixture of delta distributions centered at each element's parameter set.

*2.2.4. Evidential deep learning*

Due to the computational complexity of Bayesian methods, a number of 'single pass' methods have been proposed that seek to directly quantify the predictive uncertainty and do not require sampling. Such methods generally fall into two categories: Bayesian Last Layer techniques (Harrison et al., 2024) and Evidential Deep Learning (Han et al., 2022; Sensoy et al., 2018) or distance-based methods (Liu et al., 2023; Mukhoti et al., 2023), (Mukhoti et al., 2021a). While distance-based techniques typically rely on special regularisation approaches (e.g spectral normalisation), evidential deep learning makes no modification to the underlying network architecture, but instead models predictive uncertainty by placing a prior distribution directly over the likelihood function (Ulmer, 2021) yielding a second-order distribution that captures both epistemic and aleatoric uncertainty. In segmentation tasks, Evidental Deep Learning places a Dirchlet prior over Softmax outputs. While multiple formulations of Evidential DL exist, here we follow a framework that directly parameterises the resulting posterior distribution, and has previously been applied to medical imaging segmentation tasks producing well calibrated models with robustness to adversarial inputs (Zou et al., 2022; Li et al., 2023a). The model provides pseudo-counts $e_{vc}$ (evidence) for each class (i.e WMH or not per voxel $v$ and class $c$) and updates a uniform Dirichlet prior (i.e $\alpha_v = 1$ for Dirichlet concentration parameters $\alpha$ at voxel $v$). Finally, the posterior concentration parameters $\beta_v$ are $\beta_{vc} = (e_{vc} + 1)^2$ following (Li et al., 2023a). The Softplus function is used to ensure $e$ is positive.

To optimise the model, we calculate the Bayes risk of a chosen objective function. The corresponding Bayes risk for the cross entropy loss can be shown to equal the following (Sensoy et al., 2018):

$$\mathcal{L}_{\text{evid\_xent}} = \frac{1}{V} \sum_{v=1}^{V} \sum_{c=1}^{C} y_{vc} \left(\psi(S_v) - \psi(\alpha_{vc})\right), \tag{2}$$

where $S_v$ is the Dirichlet strength for voxel $v$, $S_v = \sum_c^C (e_{vc} + 1)^2$ where $\psi(x) = \frac{d}{dx} \log \Gamma(x)$ is the digamma function. The Bayes risk for the soft Dice loss is (Li et al., 2023a):

$$\mathcal{L}_{\text{evid\_sDice}} = 1 - \frac{2}{C} \sum_{c=1}^{C} \frac{\sum_v^V y_{vc} \frac{\alpha_{vc}}{S_v}}{\sum_v^V \left(y_{vc}^2 + (\frac{\alpha_{vc}}{S_v})^2 + \frac{\alpha_{vc}(S_v - \alpha_{vc})}{S_v^2(S_v+1)}\right)}. \tag{3}$$

To ensure that the model can represent epistemic uncertainty in unknown inputs (i.e by providing less evidence for incorrect labels and OOD inputs), misleading evidence is peanalised during training. A Kullback–Leibler (KL) divergence term between the predicted Dirchlet and a uniform Dirchlet is added to the loss for incorrectly classified voxels (Ulmer et al., 2023; Zou et al., 2022):

$$\mathcal{L}_{\text{evid\_KL}} = \sum_{v=1}^{V} \left( \log \left( \frac{\Gamma\left(\sum_{c=1}^{C} \tilde{\alpha}_{vc}\right)}{\Gamma(C) \prod_{c=1}^{C} \Gamma\left(\tilde{\alpha}_{vc}\right)} \right) \right. \\ \left. + \left( \sum_{c=1}^{C} (\tilde{\alpha}_{vc} - 1) \right) \left[ \psi\left(\tilde{\alpha}_{vc}\right) - \psi\left(\sum_{c'=1}^{C} \tilde{\alpha}_{vc'}\right) \right] \right), \tag{4}$$

where $\tilde{\alpha}_{vc} = y_{vc} + (1 - y_{vc}) \odot \alpha_{vc}$ is used to mask out correct predictions. Class probabilities for voxel $v$ are given by:

$$\frac{\beta_v}{S_v} \tag{5}$$

*2.2.5. Explicit heteroscedastic classification and stochastic segmentation networks*

An alternative approach to explicitly modelling aleatoric uncertainty is to learn a distribution over output logits. In explicit heteroscedastic classification, logits $\eta_v$ for each voxel $v$ are modelled as a Gaussian with mean $\mu(x_v; \theta)$ and variance $\sigma^2(x_v; \theta)$, with both terms predicted by our neural network. This approach may improves performance by allowing the model to control the relevance of noisy labels during training (Kendall and Gal, 2017). Output probabilities are computed by applying the Softmax function to samples drawn from this Gaussian:

$$\begin{aligned} \eta|x &\sim \mathcal{N}(\eta; \mu(x; \theta), \sigma^2(x; \theta)), \\ p(y_v|\eta) &= \text{Softmax}(\eta)_v, \end{aligned} \tag{6}$$

where $\sigma^2(x; \theta)$ is a diagonal matrix. The log-likelihood no-longer has an analytical solution, so a Monte-Carlo approximation of the loss is used:

$$\mathcal{L}_{\text{HS}} = -\text{LSE}_{s=1}^{S} \left( \sum_{i=1}^{V} \sum_{c=1}^{C} y_{ic} \log \text{Softmax}(\eta_v)_c^{(s)} \right) + \log S, \tag{7}$$

where $S$ is the number of Monte-Carlo samples, $\eta^{(s)}$ is a sample from the logit distribution and LSE is the log-sum-exponent operator.

However, treating voxels as independent cannot yield spatially coherent samples. A common approach to address this (named Stochastic Segmentation Networks (SSN)) (Monteiro et al., 2020) models the joint distribution over voxels by replacing $\sigma^2(\mathbf{x};\theta)$ with a low rank approximation of the covariance matrix $\Sigma(\mathbf{x};\theta)$: $\Sigma(\mathbf{x};\theta) = P(\mathbf{x};\theta)P(\mathbf{x};\theta)^T + D(\mathbf{x};\theta)$, where $P(\mathbf{x};\theta)$ is a $((V \times C) \times R)$ matrix, where $R$ is the rank of $P(\mathbf{x};\theta)$ and $D(\mathbf{x};\theta)$ is a diagonal matrix.

*2.2.6. Probabilistic U-Net*

The Probabilistic U-Net (Kohl et al., 2018) is an alternative approach to modelling spatial aleatoric uncertainty using a conditional variational autoencoder (cVAE) extension of U-Net. It introduces a latent space $\mathbf{z}$, conditioned on the image to encode variability in segmentations. Spatially variable segmentations are generated by sampling $\mathbf{z}$ from a learned prior network $P$ with parameters $\phi$: $P(\mathbf{z}|\mathbf{x}) = \mathcal{N}(\mu_{\text{prior}}(\mathbf{x};\phi), \sigma_{\text{prior}}(\mathbf{x};\phi))$. This latent variable is combined with the output of the last U-Net layer via a neural network $f_{\text{comb}}$ with parameters $\gamma$ which transforms $\mathbf{z}$ into meaningful spatial variation and yields the final segmentation:

$$p(\mathbf{y}|\mathbf{x},\mathcal{D}) = f_{\text{comb}}(f(\mathbf{x};\theta), \mathbf{z}; \gamma). \tag{8}$$

To ensure the latent space captures useful spatial variation, a posterior network $Q$ with parameters $\omega$ is trained, conditioned on $\mathbf{x}$ and $\mathbf{y}$: $Q(\mathbf{z}|\mathbf{x},\mathbf{y}) = \mathcal{N}(\mu_{\text{posterior}}(\mathbf{x},\mathbf{y};\omega), \sigma_{\text{posterior}}(\mathbf{x},\mathbf{y};\omega))$ The model minimises the variational lower bound (ELBO) (Eq. (9)), summing cross entropy between the model predictions with $\mathbf{z}$ generated from the posterior and the KL divergence between the prior and posterior distributions with weight $\beta$:

$$\mathcal{L}_{\text{ELBO}} = \mathbb{E}_{\mathbf{z} \sim Q(\cdot|\mathbf{x},\mathbf{y})}\left[-\log f_{\text{comb}}(f(\mathbf{x};\theta), \mathbf{z};\gamma)\right] \\ + \beta D_{\text{KL}}\left(Q(\mathbf{z}|\mathbf{y},\mathbf{x}) \parallel P(\mathbf{z}|\mathbf{x})\right). \tag{9}$$

*2.2.7. Stochastic segmentation networks ensemble*

Finally, we explore the complementary benefits of combining methods for modelling both epistemic and aleatoric uncertainty, which may improve performance (Tanno et al., 2021). Due to their simplicity and efficacy, we combine SSN with deep ensembles as an ensemble of SSN models. At inference ensemble elements are combined uniformly:

$$p(\mathbf{y}|\mathbf{x},\mathcal{D}) \approx \frac{1}{MS}\sum_{m=1}^{M}\sum_{s=1}^{S} p(\mathbf{y}|\boldsymbol{\eta}_m^{(s)}), \tag{10}$$

$$\boldsymbol{\eta}_m|\mathbf{x} \sim \mathcal{N}(\eta; \mu_m(\mathbf{x};\theta_m), \Sigma_m(\mathbf{x};\theta_m)),$$

where $\boldsymbol{\eta}_m$ and $\theta_m$ are the logits and parameters of ensemble element $m$ respectively.

*2.3. Calculating uncertainty maps*

Although some works disentangle epistemic and aleatoric uncertainty, (Charpentier et al., 2022; Laves et al., 2020), (Mukhoti et al., 2021a), achieving meaningful separation is challenging (Wimmer et al., 2023). Instead we simply assess the total (epistemic + aleatoric) uncertainty captured per technique via the predictive entropy (Eq. (11)). For stochastic (sampling based) techniques, we can view the technique as an injection of a noise vector $\mathbf{z}$ (or uncertainty index) into a deterministic model, where variations in $\mathbf{z}$ introduce variations in the segmentation.

$$\mathbb{H}\left[\mathbb{E}_{\mathbf{z} \sim p(\mathbf{z})}\left[p\left(\mathbf{y}_i|\mathbf{x}, \mathbf{z}\right)\right]\right] \approx \mathbb{H}\left[\frac{1}{S}\sum_{s=1}^{S} p\left(\mathbf{y}_i|\mathbf{x}, \mathbf{z}^{(s)}\right)\right]. \tag{11}$$

In our experiments we calculate Eq. (11) from 10 inferences for stochastic methods (Monte-Carlo Dropout, Ensembles, SSN and P-Unet), or by calculating the predictive entropy of the class probabilities for the sample free softmax entropy and Evidential DL, resulting in an image with values ranging from 0 to $\ln 2$ for all methods. For ensemble techniques we treat each element as a separate sample, while for an ensemble of SSN models we generate one sample from the logit distribution of each ensemble element.

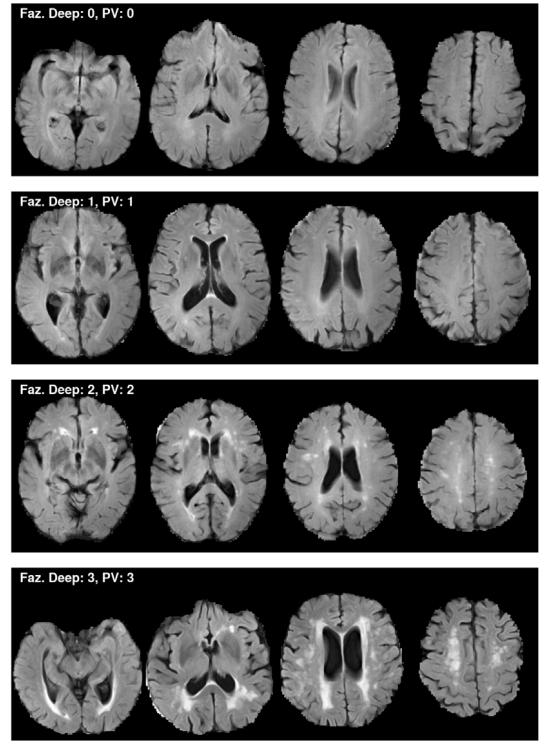

**Fig. 1.** Fazekas Categories shown where Deep WMH = PV WMH. Examples taken from the ADNI dataset.

*2.4. Fazekas estimation*

Fazekas scores are widely used by clinicians to qualitatively and visually rate WMH present in MRI images (Cedres et al., 2020) for cases of suspected vascular origin. However, WMH volume and WMH segmentation features may be used to predict the Fazekas score (Joo et al., 2022; Mu et al., 2024) automatically. Fazekas scores range from 0 to 3 in two brain regions: periventricular (PV) and deep. These regions have different correlates (Griffanti et al., 2018) and it has been argued that WMH in these regions have different etiological and histopathological substrates. WMH in the PV region exhibiting myelin loss and loosening of the white matter fibres, while in the deep these being more heterogeneous also involving ischaemic tissue damage reflected in multiple small cavities, and arteriosclerosis (Kim et al., 2008). Evidence from analysing cardiovascular risk factors suggest WMH in the deep region are more strongly linked to ischemic damage while in the PV region they have been related to age-related neurodegenerative processes. Also, some argue they are simply different stages of vascular disease that begins in the PV region and progresses to the subcortical regions (Gronewold et al., 2022), with the high correlation between total WMH volumes and Fazekas scores (Valdés Hernández et al., 2012; Joo et al., 2022) supporting this argument. We refer to deep WMH as DWMH and periventricular WMH as PVWMH. Despite their frequent usage, studies vary in their definitions of PVWMH and DWMH. However, in general PVWMH refers to WMH that is contiguous with the ventricles, and DWMH are WMH separate or sufficiently distant from the ventricles (some studies setting a threshold e.g 10 mm), referred to as the 'continuity rule' (Kim et al., 2008). Within each category, lower scores refer to no or minimal and discrete WMH, while higher scores refer to substantial areas of WMH, and confluence between PV and deep regions. Fig. 1 visualises Fazekas categories where DWMH = PVWMH. For descriptions of each category and images where DWMH $\neq$ PVWMH, see Appendix H.

Table 1
Socio-demographics, clinical characteristics, visual Fazekas scores (Fazekas et al., 1987) and experimental usage of each of the datasets.

| Characteristic | CVD Dataset | ADNI Dataset | WMH Challenge | MSS3 Dataset |
| --- | --- | --- | --- | --- |
| **Participants (N)** | 250 | 298 | 60 | 65 |
| **Females (N)** | 88 | 136 | – | 16 |
| **Mean Age (SD, years)** | 71 (12) | 72 (7.3) | – | 67.6 (19.7) |
| **Hypertension (%)** | 70% | 14% | – | – |
| **No. of MRI Protocols** | 4 | 13 | 3 | 1 |
| **Fazekas Scores** | Deep: 1 [1, 2] | Deep: 1 [0, 1] | Deep: 2 [1, 2.5] | Deep: 1 [1, 2] |
| **(median [Q1, Q3])** | PV: 2 [1, 2] | PV: 1 [1, 1] | PV: 2 [1, 3] | PV: 1 [1, 2.5] |
| **Experimental Usage** | | | | |
| UQ methods comparison: | | | | |
| Train (Test) | ✓ (–) | – (–) | – (✓) | – (–) |
| Fazekas prediction: Train (Test) | ✓ (✓) | ✓ (✓) | ✓ (✓) | – (✓) |
| Quality control: Train (Test) | ✓ (✓) | – (–) | ✓ (✓) | – (–) |
| Qualitative analysis | – | ✓ | ✓ | ✓ |

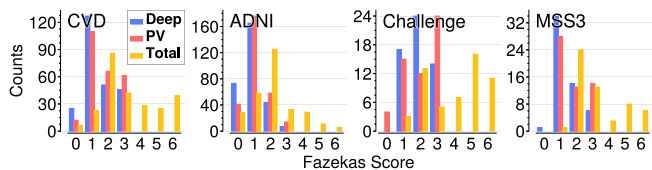

**Fig. 2.** Distribution of Fazekas score per dataset, divided into the Deep, Periventriclar (PV) and Total (combined) regions for each of the four datasets. Datasets appear in the following order: CVD, ADNI, Challenge and MSS3 from left to right.

## 3. Materials and methods

### 3.1. Data

We use T1-weighted and Fluid Attenuated Inversion Recovery (FLAIR) brain MRI data from four sources: (1) the cerebrovascular disease (CVD) dataset, (2) the Alzheimer's disease neuroimaging initiative (ADNI), (3) the 2018 WMH segmentation MICCAI challenge (Challenge) and (4) the Mild Stroke Study 3 (MSS3). Table 1 summarises the data, while Fig. 2 shows the distribution of the degree of WMH severity, given by Fazekas scores, for the four datasets. We use the images from the CVD dataset to train the UQ techniques for WMH segmentation. We use the Challenge dataset for evaluating the performance of the UQ techniques. We use the MSS3 and ADNI datasets for (1) evaluating the robustness of UQ schemes to unseen data with different disease and population characteristics, and (2) assessing the usefulness of UQ schemes in the Fazekas classification task. The CVD and Challenge datasets are further used in training and testing for the Fazekas and Quality Control tasks. Further details are in Section 3.3, 3.5, 3.6. MRI acquisition protocols for all data used in this work are described in Appendix B. WMH reference annotations were provided for the CVD and Challenge datasets. Fazekas visual scores, generated by an experienced neuroradiologist, were provided for the CVD and MSS3 datasets. Fazekas scores for Challenge and ADNI datasets are not provided as part of the datasets. They were generated by an expert rater for this study. The CVD and MSS3 data are not publicly available but can be made available by request to the primary studies' data holders.

#### 3.1.1. CVD dataset

The CVD dataset consists of data from 250 participants (88 females) in ageing and mild ischaemic stroke studies (Guerrero et al., 2018) at the Centre for Clinical Brain Sciences at the University of Edinburgh. Mean (SD) age of the cohort is 71 (12), with the following distribution of vascular risk factors: 30 (12%) have diabetes, 174 (70%) are hypertensive, 139 (56%) have high cholesterol, and 145 (58%) are ex- or current smokers. In addition to WMH, 92 (37%) scans present one stroke lesion cluster and 66 (26%) present two. All images were acquired at a 1.5T GE Signa HDxt scanner, but by their acquisition protocols they can be grouped in 4 different unbalanced domains.

#### 3.1.2. ADNI dataset

The Alzheimer's Disease Neuroimaging Initiative (ADNI) database (adni.loni.usc.edu) was launched in 2003 as a public–private partnership, led by Principal Investigator Michael W. Weiner, MD. The primary goal of ADNI has been to test whether serial magnetic resonance imaging (MRI), positron emission tomography (PET), other biological markers, and clinical and neuropsychological assessment can be combined to measure the progression of mild cognitive impairment (MCI) and early Alzheimer's disease (AD). We utilise a subset of this dataset, which consists of structural brain MRI from 298 ADNI participants (136 females). Unlike the CVD and WMH segmentation challenge datasets, for which the data were deliberately selected from clinical studies to represent the breadth of WM disease found in individuals with sporadic SVD and the elderly, this dataset included data from all ADNI participants that in January 2016 had three consecutive MRI taken 12 months apart, with not just the T1-weighted MRI sequence, but also the rest of the structural sequences that allow for detection of SVD features. The data were extracted and screened for quality for previously published analyses (Valdés Hernández et al., 2018; Harper et al., 2018). This subsample's mean age (SD) is 72 (7.3) years old. Only 43 (14%) are hypertensive and 126 (42%) have endocrino-metabolic risk factors for Alzheimer's disease (e.g diabetes, hypertension, hyperlipidemia, and obesity). In general, 249 (84%) had one or more cardiovascular risk factors. From the 59 ADNI imaging centers, 37 contributed with images to this subsample, which can be grouped in 13 unbalanced domains acquired from multiple MRI scanner models. Due to the lack of WMH reference segmentations available for the ADNI dataset, we cannot directly assess the generalisability and robustness of our WMH segmentation models on this dataset. However, we use the ADNI dataset for qualitative analysis and for the Fazekas prediction task, as a way to validate our model outputs on a data distribution which is unseen during training the WMH segmentation models.

#### 3.1.3. WMH challenge dataset

This dataset belongs to the publicly available MICCAI WMH Segmentation Challenge (Kuijf et al., 2019). It consists of T1-weighted, FLAIR, and WMH ground-truth images from 60 patients, 20 of who were imaged from 3 separate sites with different acquisition parameters.

#### 3.1.4. MSS3 dataset

MSS3 (Clancy et al., 2021) is a prospective observational cohort study to identify risk factors for and clinical implications of SVD progression. We use a subsample of 65 patients (16 females) aged 67.6 (19.7) years, extracted as part of a previous analysis (Philps et al., 2024). Explanation of the segmentation pipeline can be found at Valdés Hernández et al. (2023).

## 3.2. Data preprocessing

Brain extraction was performed using ROBEX (Iglesias et al., 2011) to all datasets except CVD where the intracranial volume binary masks provided were used. Correction from b1 magnetic field inhomogeneities was performed to the T1-weighted images from all datasets using FSL-FAST (Zhang et al., 2001). Given the multi-domain nature of all datasets used, we resampled all images to 1 x 1 x 3 mm$^3$ voxel size (i.e., the median voxel size of the sample) using 3rd-order B-spline interpolation. WMH ground-truth masks from the CVD and WMH segmentation Challenge datasets were resampled using nearest neighbour interpolation. We aligned both MRI sequences in the ADNI dataset using Simple-ITK (Lowekamp et al., 2013) (all other images were spatially aligned when these datasets were accessed). Z-score intensity normalisation (Reinhold et al., 2019) was applied to each scan sequence per patient.

Supplementary A.1 shows the voxel intensities for the CVD, ADNI and Challenge datasets pre and post z-score normalisation. After normalisation, we see that the histograms of the FLAIR intensities in each dataset and subdomain can be approximated as a bimodal distribution, with a large peak centered slightly right of 0 and a lower peak below zero. The relation between the histograms' peak of each mode differs between the datasets, with the WMH Challenge dataset showing the clearest distinction between the two modes, while the CVD and ADNI datasets peak mainly around the positive mode (i.e closer to a unimodal negative skewed distribution). Along with the variations in cohort selection and Fazekas distributions these datasets represent real world differences in input distribution through which we shall assess the generalisation performance and utility of the examined UQ techniques.

## 3.3. Backbone segmentation method

We initially assess each UQ technique using a U-Net segmentation architecture with residual connections in each block adapted from Mojiri Forooshani et al. (2022). The U-Net architecture is an effective and parameter efficient architecture in many segmentation tasks (Isensee et al., 2021). To increase the effective sample size of the training data, we use augmentation procedures similar to previous work on the WMH challenge dataset (Park et al., 2021), including random flipping in the sagittal plane (p=0.5), affine transformations (each p = 0.2) such as rotation (−15°, 15°), shearing (−18°, 18°), translation with ratio (−0.1, 0.1). Segmentation models are trained using the CVD dataset and evaluated using the Challenge dataset to assess OOD generalisation performance. In order obtain segmentation and uncertainty map outputs for all images in the CVD dataset (used in later analysis (see Section 3.5), we employ 6 fold cross validation. For the purpose of hyperparameter tuning and early stopping we further split each training fold to obtain a 70/15/15% train, validate, test split. Hyperparameter tuning is performed using the first split only. Splits are calculated on 3D images, after which we use 2D axial slices for training the model. All evaluation is performed in 3D. For UQ methods that express uncertainty via producing different inferences (samples) for the same input, to generate 3D samples, we order the samples over each 2D slice by volume, such that sample $n$ contains the $n$th largest sample of each slice. This promotes diversity of sample volume and allows us to estimate the range of WMH volumes over the whole brain implied by the model. We provide our code here[2].

[2] https://github.com/BenjaminPhi5/wmhuq

## 3.4. Experiment 1: Comparison of UQ methods

### 3.4.1. Implementation of UQ methods

We begin with training a baseline deterministic segmentation model, trained with the combo loss (Taghanaki et al., 2019), which is a sum of the cross entropy and soft Dice score loss (we use an equal weighting). We reuse combo loss or an equivalent for consistent comparison between methods (as outlined below for each method). The choice of loss function can effect not just the calibration of a given model but the utility of the resulting uncertainty map (Philps et al., 2023). Since many axial slices contain no WMH, we tuned the proportion of empty slices that are retained in the training data, selecting from the set $0.05, 0.1, 0.3, 0.5, 0.9, 1$. We found that a retention proportion of 0.1 was optimal. For each UQ technique, we separately tuned the learning rate in the set $0.01, 0.001, 0.0001$ and weighted decay in between $[0, 1e-2]$ to prevent overfitting. We multiplied the learning rate by 0.5 every 50 epochs until the learning rate reaches $2e-5$. Training stops if there has been no improvement in the validation loss for 100 epochs. We use a batch size of 32. Below we outline any extra loss function configuration extra hyperparameter tuning applied to each method on the validation data, after selecting the parameters that yielded the strongest performance on the majority of the following metrics [Dice, Top (Highest) inference Dice, AVD, Top (lowest) inference AVD] unless otherwise specified.

**MC-Dropout (MC-Drop):** We tuned the dropout rate, from [0.05, 0.1, 0.2, 0.3 0.4] applied to {encoder only, all, or decoder only} layers. We found that using all with a rate of 0.2, yields the highest sample diversity and the strongest Dice performance. Applying dropout to only part of the network architecture significantly reduced the sample diversity and could lead to poor performance on metrics such as $D^2_{\text{GED}}$ (see 3.4.3). Due to the regularising effect of Dropout, weight decay was not necessary.

**Deep Ensembles (Ens):** We initially adjusted the number of elements in the ensemble in the range [3, 5, 10, 20, 30], finding that 10 presents a suitable balance between performance and training time. Using a smaller ensemble significantly reduced the sample diversity expressed by the ensemble, with diminishing returns in increasing the ensemble size beyond 10.

**Evidential Deep Learning (Evid):** To retain the combo loss for the evidential DL UQ scheme, use the sum of Eqs. (3), (2), (4). We further tune a weighting parameter for the KL component (Eq. (4)) from the set $0, 0.01, 0.05, 0.1, 0.5, 1$ finding 0.05 yields the strongest performance. We initially adjusted the value of the KL term in the loss on a logarithmic scale between 0 and 10. For stability during training, we increase the KL term from 0 to its target value over the first 50 epochs.

**Stochastic Segmentation Networks (SSN):** We calculate the combo loss as the sum of Eq. (7) with the soft dice loss applied to the mean and inferences generated from the model: $\mathcal{L} = \mathcal{L}_{\text{HS}} + 0.5\mathcal{L}_{\text{SDC}}(\text{Softmax}(\mu(\mathbf{x};\theta)), \mathbf{y}) + \frac{0.5}{S}\sum_{s=1}^{S}\mathcal{L}_{\text{SDC}}(\text{Softmax}(\eta^{(s)}), \mathbf{y})$, where $\mathcal{L}_{\text{SDC}}(a, b)$ is the soft Dice loss between vectors $a$ and $b$. We initially adjusted the rank of the covariance matrix, trying values of [5, 10, 12, 15, 18, 20, 25, 30, 40, 50] with 25 yielding strong performance. Lower ranks reduced the expressiveness of the sample distribution (i.e highest Dice amongst multiple inferences) while higher ranks produced marginal to non-existent gains. We also train a model with rank of 1 (diagonal covariance) for the variant where voxels are treated as independent (Eq. (6)) - we refer to this variant as **(Ind)**. We refer to an ensemble of 10 SSN models with covariance rank of 25 as **(SSN Ens)**.

**Probabilistic UNet (P-Unet):** We initially adjusted the KL term between 0 and 10 on a logarithmic scale, finding 1 most effective. We tune the size of the latent dimension $z$ (see Section 2.2.6) in $6, 12, 24, 36$ finding 12 to as effective as 24 or 36. To compute the combo loss, we add the soft Dice component to the ELBO (Eq. (9)) using $z$ generated from the posterior network. We employed the existing implementation of P-Unet (Kohl et al., 2018), and modified it to integrate it with our segmentation backbone.

### 3.4.2. Comparison with nnunet backbone

To compare the utility and semantic properties of UQ techniques across different training paradigms, we further implement the baseline SEnt and the SSN and SSN-Ens UQ methods, chosen due to their segmentation and UQ map performance, using the popular and often state-of-the-art nnUNet segmentation model and training paradigm (Isensee et al., 2021). To implement a SSN variant of nnUNet, we modify the backbone architecture recommended by nnUNet by adding a SSN output head to each decoder stage. During training, we then apply the SSN variant of the combo loss to each decoder head, mirroring the deep supervision setup used by nnUNet, while keeping all other nnUNet settings the same (e.g., augmentation procedure, optimisation parameters).

### 3.4.3. Evaluation metrics

This section outlines the metrics used to assess the quality of the segmentations and uncertainty maps generated from applying the different UQ techniques evaluated. The WMH segmentation is considered as the voxelwise WMH probability map thresholded at 0.5 (the mean WMH probability map is used for stochastic models). For each metric $E$, errorbars in figures report the standard deviation (with Bessel's correction) across the 6 model runs (of the mean metric score over the subjects):

$$\text{std}\left(\frac{1}{S}\sum_{s=1}^{S} E_{;s}\right), \tag{12}$$

while tables further report the standard deviation across subjects (of the mean score over model runs):

$$\text{std}\left(\frac{1}{R}\sum_{r=1}^{R} E_{r;}\right), \tag{13}$$

where $E_{rs}$ is metric value for model run r and subject s.

**Segmentation Metrics:** Before assessing the quality of the uncertainty quantification provided by each technique, we first assess the segmentation quality of each technique. Accurate overall WMH volume estimation is essential for WMH segmentation methods to be useful in research and clinical settings, along with the identification of individual WMH components. Hence we use the Absolute Volume Difference (AVD) and per-cluster F1 score, in addition to the Dice score, in order to assess segmentation quality. Due to the large size disparity between individual WMH, with many very small WMH often present in an image, we count WMH as detected if the intersection over union (IOU) is $> 0$.

**Top Sample and Generalised Energy Distance:** Assessing pixelwise calibration fails to capture structural variance expressed by a segmentation model (i.e differences in boundary shape). A method that produces a meaningful distribution of WMH segmentations should ideally contain a sample very close to the ground truth within a small number of samples. This for example may allow a user to select an output from a range of possible segmentations. For techniques that generate multiple inferences (samples) per input $x$, we report the highest scoring inference on the Dice and AVD metrics - we refer to this as the 'Top' score. We further compute the generalised energy distance (Székely and Rizzo, 2013), which measures the distance between the predicted distribution and the ground truth distribution and a commonly used metric for assessing sample quality (Czolbe et al., 2021; Monteiro et al., 2020): $D^2_{\text{GED}}(p, \hat{p}) = 2\mathbb{E}_{y \sim p, \hat{y} \sim \hat{p}}[d(y, \hat{y})] - \mathbb{E}_{y, y' \sim p}[d(y, y')] - \mathbb{E}_{\hat{y}, \hat{y}' \sim \hat{p}}[d(\hat{y}, \hat{y}')]$, where $d$ is a distance metric, e.g $d(a, b) = 1 - \text{IOU}(a, b)$. 10 inferences are generated for each technique for calculating both Top sample and Generalised Energy Distance metrics.

**sUEO Score:** Uncertainty maps should precisely identify regions with errors. The uncertainty error overlap (UEO) metric assesses overlap between uncertainty and segmentation error by computing the Dice metric between a thresholded uncertainty map and the error in the mean segmentation (Jungo et al., 2020). This rewards uncertainty that is well localised to the segmentation error. Soft Uncertainty-Error Overlap (sUEO) (Li et al., 2023a), uses a soft Dice score to provide a threshold free evaluation of overlap between uncertainty and overlap. sUEO is defined as:

$$\text{sUEO} = \frac{2\sum_i e_i u_i}{\sum_i e_i^2 + u_i^2}, \tag{14}$$

where $e_i$ is the segmentation error at pixel $i$: $e_i = \text{argmax}_c p_{ic} \neq y_i$. Higher sUEO denotes greater overlap.

**Patchwise Uncertainty quality metrics:** The sUEO metric assess the uncertainty at an image level. To assess the uncertainty map at subregions within the image, we use the patchwise uncertainty quality metrics proposed in prior work (Mukhoti and Gal, 2019b). Ideally, our model should be accurate when it is certain and by extension uncertain when it is inaccurate. The probability that a model is accurate when it is certain, $p(acc|cert)$, and the probability that a model is uncertain when it is inaccurate, $p(uncert|inacc)$, are calculated by counting regions in the image where the model is accurate or inaccurate, and certain or uncertain respectively. To do this, we divide the image into patches of size $4^3$. For each patch, it is accurate if the proportion of accurately segmented voxels is $\geq 0.8$. We plot how these metrics change as we change the uncertainty threshold $\tau$ (patches are uncertain if the average uncertainty in the patch is $\geq \tau$). Finally, the PAvPU (Patch Accuracy vs. Patch Uncertainty) metric (Mukhoti and Gal, 2019b) combines these two metrics:

$$\text{PAvPU} = \frac{n_{ac} + n_{ui}}{n_{ac} + n_{ui} + n_{au} + n_{ci}}, \tag{15}$$

where $n_{ac}$ refers to the number of patches that are accurate and certain, $n_{ui}$ for patches that are uncertain and inaccurate, $n_{au}$ for patches that are accurate and uncertain, and $n_{ci}$ for patches that are certain and inaccurate.

**Missing Lesion Coverage:** WMH can be very small, particularly for cases with thin pencil-like periventricular lesions and deep isolated WMH. However aleatoric uncertainty is high for such lesions due to their relatively high proportion of edge voxels to total voxels and their subjective nature. Furthermore, small lesions can be easily missed by the model, which can result in silent failure, where the lesion is missed in the segmentation, and not highlighted as uncertain either. To assess the coverage of false negatives in the uncertainty map and lesion-wise silent failure rates, we calculate three metrics. First, we calculate the false negative coverage (per voxel) in the uncertainty map. Second, we calculate the proportion of instances that are undetected (missed) in both the mean predicted segmentation and the uncertainty map. We define instances as 3D connected components in a binarized image, and an undetected instance as an instance where no voxels in the segmentation or thresholded uncertainty map overlap with the instance. Similarly, we define detected instances where at least one voxel in the predicted segmentation or thresholded uncertainty map does overlap with the instance. Finally, we compute the size of missed instances.

### 3.5. Experiment 2: Downstream prediction of the fazekas score

In this subsection we outline our methodology for utilising WMH segmentation probability maps and uncertainty maps to predict the clinically useful Fazekas score, across multiple datasets with different cohort and disease characteristics. We utilise outputs from our WMH segmentation models trained on the CVD dataset, extracting spatial and volumetric features (outlined below) from the outputs which we use to train a logistic regression classifier. In order to predict Fazekas, we use the extracted features to fit a logistic regression model. In order to assess whether the choice of segmentation model impacts classification performance, we shall extract features from the SEnt model, as well as the SSN-Ens and P-Unet UQ techniques.

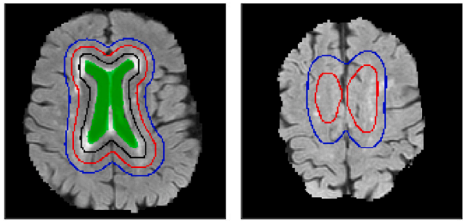

**Fig. 3.** Ventricle rings used in the feature extraction stage for the Fazekas/QC prediction tasks. Green shows the ventricle segmentation. Black, Red and Blue lines show euclidean distances of 5, 10 and 15 mm from the ventricles respectively. Example taken from the Challenge dataset.

*3.5.1. Feature extraction rationale and thresholding*

In this section we outline the features extracted for the prediction of the Fazekas score. Given that WMH volume is a known predictor of both DWMH and PVWMH Fazekas score (Andere et al., 2022), we calculate the sum of voxel intensities in the thresholded maps. However, as Fazekas is also defined by the spatial position and confluence of WMH, we devise a feature extraction technique to capture all these components. First, we generate rings around the ventricles, at a distance of < 5 mm, between 5 mm and < 10 mm, between 10 mm and < 15 mm and finally ≥ 15 mm away from the ventricles (Fig. 3). Synth-Seg (Billot et al., 2023) is used for ventricle segmentation. For each region, we extract the total volume, the number of connected components per volume, and the standard deviation of connected components per volume. We further extract the largest connected component that includes both the 1st and 2nd ring as well as the 2nd and 3rd ring, in order to assess confluence. To assess the utility of the uncertainty information in improving the Fazekas score, we also extract the same features from the uncertainty map of each method. We further extract the standard deviation of each feature across 10 inferences generated from sample-based UQ methods.

*3.5.2. Model training and hyperparameter tuning*

There are two key hyperparameters in our method for predicting Fazekas scores. The first is the threshold at which we filter out information from the segmentation probability map (the mean WMH probability map for stochastic models) and uncertainty maps, prior to calculating the features from the previous section. For a given threshold $t$ we set all voxels with an intensity below $t$ to 0. We apply the same threshold to the UQ and segmentation map. The second is the total number of possible features we could provide to the model, which is large. To prevent overfitting with our relatively small sample size, we employ recursive feature elimination, where we train a model on the feature set multiple times, each time removing the least important features (by assessing their importance weighting) until we have $k$ features remaining. We report results for $t = 0.2$ and $k = 18$ but to explore the impact of hyperparameter settings on performance, we further run the experiment for settings of $t$ in $0.1, 0.2, 0.3$ and $k$ in $[10, 26]$, and report the results across all possible hyper-parameter settings.

All features are z-score normalised, with the mean and standard deviation calculated excluding values above the 95th percentile to reduce the impact of outliers. Values above the 95th percentile are clipped to the 95th percentile prior to normalisation. We use a multivariable logistic regression model for predicting Fazekas score, with a balanced cross-entropy loss that reweights the loss of each class (i.e., score) inversely proportional to the number of elements of that class. While reweighting the loss may reduce the overall accuracy of the model across the entire cohort, achieving greater separation between Fazekas scores of 0,1 and of 1,2 (Fazekas 1 is the dominant class) is arguably more useful given our highly unbalanced and not generally representative sample. This allows us to better detect patients who have less visible white matter disease (Fazekas 0) and patients who already demonstrate more severe WMH (Fazekas 2 and 3), which has been associated with cognitive dysfunction, and it is useful for identifying patients with increased risk of cerebrovascular disease (Cedres et al., 2020; Wardlaw et al., 2013). We use a regularisation strength (weight decay) of 10 to prevent model overfitting. We combine the CVD, ADNI and Challenge datasets, with a 75%/25% train test split, and leave the MSS3 as a separate held out test set. For the CVD dataset, we only extract predictions and UQ maps from models for which the given image was in the test fold during the original model training. For all other datasets, we used the models trained from the first cross validation fold. We train separate classifiers for the Deep and PV categories.

*3.5.3. Classification metrics*

We employ the Cohen's Kappa agreement coefficient, balanced accuracy score and the area under the Receiver Operating Characteristic curve (AUROC) score to assess classification performance. To assess the calibration of each method, we report the root brier score: $\sqrt{\frac{1}{N} \sum_{n=1}^{N} \sum_{c=1}^{C} \left( p_c^{(n)} - y_c^{(n)} \right)^2}$ where $N$ is the number of test time instances, $C$ is the number of classes and $p$ and $y$ are the model predictions and ground truth labels respectively. We report the mean and 95% confidence intervals, obtained via bootstrapping (percentile method) using 1000 randomly sampled train/test splits to train the model.

*3.6. Experiment 3: Quality control of the segmentation*

We wish to explore the usefulness of different methods for performing quality control (QC), identifying poor quality segmentations (hereon referred to as the QC task). We setup a binary classification task, defining images with Dice ≤ 0.57 as 'poor quality' (this corresponds to images in the bottom 20th percentile in the CVD test set) and all other images are 'good quality'. We use the same feature extraction and model training process as for the Fazekas classification task. The CVD test set and Challenge dataset are combined and divided into 75%/25% train test split. We use the performance of the SEnt model to derive the class labels. Due to the smaller number of images available during training than those available for the Fazekas scoring task, we report results for $t = 0.2$, $k = 9$ and restrict the tested range of $k$ to $[6, 12]$. We employ the same evaluation metrics as Section 3.5.3.

*3.7. Experiment 4: Qualitative analysis*

To understand the semantic properties of UQ maps under different conditions, we perform a qualitative analysis of the UQ maps, visually comparing the output from the SSN-Ens and baseline SEnt methods for both the UNet standard model and nnUNet variants. We select examples from the Challenge dataset where both SSN-Ens and SEnt models fail to segment small WMH (WMH partially or entirely unsegmented in the WMH segmentation map). To examine how the methods perform under conditions with confounding information in the image, we further select images of cortical and sub-cortical stroke lesions of varying sizes from the MSS3 dataset and images with image artefacts from the ADNI dataset. Uncertainty maps are visualised without thresholding to compare regions of low and high uncertainty.

## 4. Results

*4.1. UQ techniques improve model robustness and reduce silent failure rates in UQ maps*

First we inspect the segmentation performance of each method on the OOD (Challenge) dataset. Table 2 shows the segmentation performance metrics. All methods outperform the baseline SEnt Dice score of for the standard models, except for the nnUNet variant of SSN, while nnUnet show higher Dice scores and per-component F1 scores overall. However, the overall WMH volume estimation is poorer for the nnUnet variants due to increased overestimation of the volume, with the SEnt yielding a poorer AVD% score (51.2%) than the baseline SEnt model (50.2%). While SSN-Ens again improves the AVD% score to 45.6%, results are still poorer than the baseline SSN-Ens model at 42.7%. Methods that model epistemic uncertainty always yield an improved AVD% score over the baseline, with ensembling techniques the most effective. Modelling both uncertainty components provides complementary performance benefits, with SSN-Ens yielding higher precision over recall (Fig. 4) and Dice (0.70 for standard and 0.72 for nnUNet models) over modelling just aleatoric (SSN - Dice 0.68 for standard and 0.70 for nnUNet) and epistemic (Ens - Dice 0.69 for standard model) vs baseline SEnt (Dice 0.67 for standard and 0.7 for nnUNet) as well as yielding the highest Top Dice and lowest Top AVD% for standard and nnUnet models. MC-Dropout also yields meaningful variation between model inferences, improving AVD% from 46.5% for the mean prediction to 23.9% for the best sample, suggesting that Dropout layers early in the model contribute to substantial structural changes in the segmentation. Nonetheless, This is considerably higher than the Top AVD% scores achieved by SSN-Ens of 12.3% and 11.0% for the standard and nnUnet models respectively. Unsurprisingly, Ind is unable to express meaningful differences in the quality of segmentation with minimal differences in the Top vs mean Dice and AVD% in both datasets. The sample diversity of P-Unet is also poor, with limited improvement in Top AVD% over AVD%. The SSN-Ens model is more consistent across model runs, with less variability across both model runs and subjects than all other methods except for Ens on the AVD% score (Table 2). All methods outperform the baseline in detection of individual WMH, with combining Ens and SSN again providing greater improvements than either technique individually.

Table 2
Segmentation performance for each technique on the OOD (Challenge) dataset. Top Dice/AVD refer to the best score attained from the 10 samples (forward inferences) generated from each model. The mean and standard deviation (Eq. (12)) were obtained from the 6 model runs, while the standard deviation of each metric across all subjects is reported in parentheses (Eq. (13))

|  | Dice | Top Dice | AVD% | Top AVD% | Per-component F1 |
| --- | --- | --- | --- | --- | --- |
| SEnt | 0.67 ± 0.006 (0.16) | — | 50.2 ± 6.4 (110.1) | — | 0.47 ± 0.018 (0.10) |
| MC-Drop | 0.68 ± 0.003 (0.16) | 0.70 ± 0.004 (0.15) | 46.5 ± 4.6 (96.5) | 23.9 ± 4.0 (61.6) | 0.50 ± 0.004 (0.11) |
| Ens | 0.69 ± 0.002 (0.11) | 0.69 ± 0.004 (0.11) | **42.7 ± 1.7 (69.9)** | 19.2 ± 2.5 (39.2) | 0.51 ± 0.003 (0.09) |
| Evid | 0.68 ± 0.008 (0.16) | — | 49.2 ± 7.6 (94.5) | — | 0.52 ± 0.010 (0.10) |
| Ind | 0.69 ± 0.009 (0.14) | 0.69 ± 0.008 (0.14) | 50.1 ± 7.5 (89.3) | 48.1 ± 7.4 (87.7) | 0.53 ± 0.014 (0.10) |
| P-Unet | 0.68 ± 0.013 (0.16) | 0.69 ± 0.011 (0.15) | 55.4 ± 12.4 (92.2) | 39.2 ± 8.4 (73.2) | 0.54 ± 0.024 (0.10) |
| SSN | 0.68 ± 0.013 (0.13) | 0.71 ± 0.012 (0.12) | 55.0 ± 10.9 (75.7) | 17.1 ± 5.4 (36.5) | 0.53 ± 0.015 (0.09) |
| SSN-Ens | **0.70 ± 0.004 (0.11)** | **0.72 ± 0.004 (0.09)** | 43.8 ± 2.9 (59.4) | **12.3 ± 1.4 (27.8)** | **0.57 ± 0.003 (0.09)** |
| SEnt (nnUNet) | 0.70 ± 0.012 (0.15) | — | 51.2 ± 9.9 (95.4) | — | 0.59 ± 0.016 (0.10) |
| SSN (nnUNet) | 0.70 ± 0.009 (0.14) | 0.74 ± 0.005 (0.12) | 49.0 ± 11.0 (86.7) | 13.4 ± 4.4 (34.0) | 0.61 ± 0.022 (0.10) |
| SSN-Ens (nnUNet) | **0.72 ± 0.006 (0.14)** | **0.74 ± 0.003 (0.12)** | 45.6 ± 6.1 (85.9) | **11.0 ± 2.2 (32.2)** | **0.63 ± 0.009 (0.10)** |

Table 3
Uncertainty quality for each technique on the OOD (Challenge) dataset. sUEO: Soft Uncertainty Error Overlap. $D^2_{\text{GED}}$: Generalised energy distance. The mean and standard deviation (Eq. (12)) were obtained from the 6 model runs, while the standard deviation of each metric across all subjects is reported in parentheses (Eq. (13))

|  | sUEO | $D^2_{\text{GED}}$ |
| --- | --- | --- |
| SEnt | 0.47 ± 0.004 (0.04) | — |
| MC-Drop | 0.47 ± 0.007 (0.05) | 0.73 ± 0.005 (0.26) |
| Ens | 0.46 ± 0.003 (0.04) | 0.66 ± 0.005 (0.17) |
| Evid | 0.48 ± 0.004 (0.03) | — |
| Ind | 0.48 ± 0.007 (0.03) | 0.89 ± 0.020 (0.30) |
| P-Unet | **0.50 ± 0.011 (0.03)** | 0.87 ± 0.024 (0.32) |
| SSN | 0.46 ± 0.004 (0.03) | 0.71 ± 0.032 (0.23) |
| SSN-Ens | 0.46 ± 0.002 (0.03) | **0.63 ± 0.008 (0.17)** |
| SEnt (nnUNet) | 0.47 ± 0.011 (0.04) | — |
| SSN (nnUNet) | 0.47 ± 0.011 (0.05) | 0.66 ± 0.025 (0.24) |
| SSN-Ens (nnUNet) | 0.45 ± 0.005 (0.05) | **0.61 ± 0.014 (0.23)** |

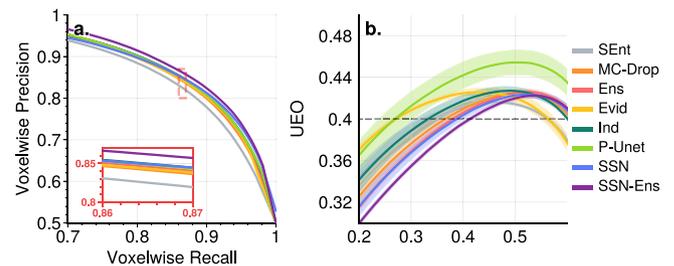

Fig. 4. Voxelwise precision vs recall for WMH classification in the OOD (Challenge) dataset. Red box shows zoomed in view of the dashed region to demonstrate the ranking of method performance. (b) Mean Uncertainty error overlap (UEO) as $\tau$ changes for the OOD (Challenge) dataset. Dotted line shows UEO at 0.4. Shaded area shows the standard deviation across the 6 model runs (Eq. (12)).

Table 3 demonstrates the uncertainty quality scores. P-Unet yields the strongest sUEO (0.5) score while methods that perform the strongest on the Top AVD% metric (i.e Ens, SSN, SSN-Ens) yield equal or lower sUEO score than the baseline 0.47. We see this is opposite to the $D^2_{\text{GED}}$ where SSN-Ens yields the best (lowest) score (0.63 for standard and 0.61 for nnUnet models) while P-Unet (0.87) and Ind (0.89) perform very poorly, as methods with high sample diversity achieve lower $D_{\text{GED}}^2$ (Fig. 8).

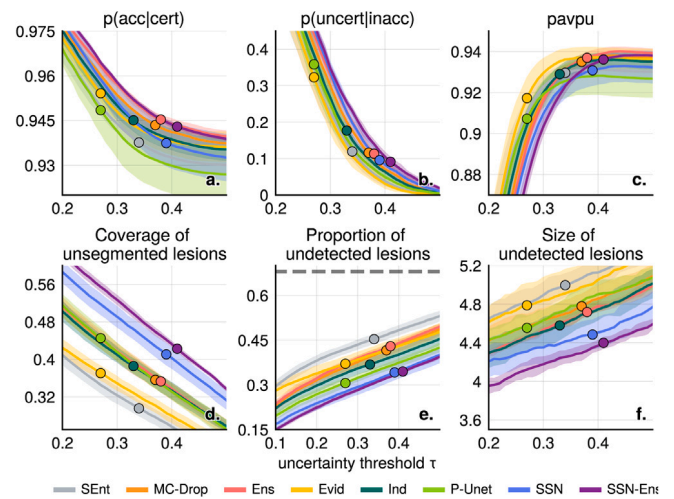

Fig. 5. (a-c) Patchwise uncertainty metrics as $\tau$ changes for the OOD (Challenge) dataset. All metrics computed with a sliding window size of $4^3$ voxels and accuracy threshold of 0.8. (a) Probability that a patch is accurate given that it is certain. (b) Probability that a patch is uncertain given that it is inaccurate. (c) PAvPU metric. (d-f) Lesion Instance Coverage metrics for the OOD (Challenge) dataset. Instances are 3D connected components in the ground truth WMH mask. (d) Coverage of unsegmented lesions: The mean proportion of unsegmented instances which are deemed uncertain. (e) Proportion of undetected lesions: The proportion of unsegmented instances for which less than 50% of the instance (or less than 5 voxels, whichever is lower) are deemed uncertain. Dashed line shows the number of instances which are undetected for baseline model SEnt in the WMH segmentation. Most instances are < 10 voxels in volume and undetected. (f) The size of undetected instances. For all plots, the shaded area denotes the standard deviation over the 6 model runs (Eq. (12)), and the circles indicate the point at which each method attains an Uncertainty Error Overlap (UEO) of 0.4 on the OOD (Challenge) dataset.

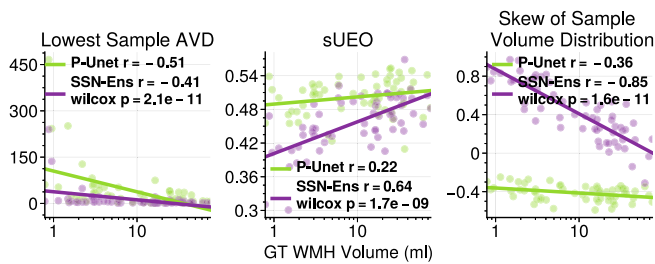

**Fig. 6.** Comparison of P-Unet and SSN for each subject in the OOD (Challenge) dataset, reporting Pearson correlation coefficient and Wilcox signed-rank test. Lowest Sample AVD score: SSN-Ens exhibits consistently lower (better) top sample AVD score ($p = 2.1e-11$) sUEO score: P-Unet exhibits consistently higher (better) soft Uncertainty Error Overlap (sUEO) score than SSN-Ens ($p = 1.7e-9$).

While sUEO rewards strong localisation of errors, this does not differentiate between uncertainty around the boundaries of large WMH and the coverage of deep isolated WMH, which are proportionally insignificant but important indicators of lesion growth. sUEO hence inherently favours models that introduce less uncertainty around correctly identified voxels at the cost of detecting small WMH. Fig. 5 (top) shows the patchwise uncertainty quality and lesion coverage metrics. SSN-Ens yields the highest score for $p(acc|cert)$ and $p(uncert|inacc)$ across all thresholds, while P-Unet performs the strongest and SSN-Ens the weakest on the PAvPU metric (due to the high number of uncertain but accurate voxels and hence low UEO score at low $\tau$ (Fig. 4)). When adjusted for equivalent UEO scores, SSN-Ens yields now the poorest performance on $p(acc|cert)$ and $p(uncert|inacc)$ metrics (while the strongest performance on the combined PAvPU metric), due to the high threshold needed to reduce the number of uncertain accurate voxels.

Fig. 5 (bottom) shows the results for the detection of unsegmented WMH lesions. While SSN-Ens yields consistently the highest coverage of unsegmented WMH regardless of $\tau$ (a), when adjusted for UEO it is comparable to P-Unet. SEnt is the worst performer with the lowest coverage proportion, and highest proportion and size of undetected lesions, with the gap between the UQ techniques and SEnt increasing once adjusted for equivalent UEO scores. Hence, the patch-wise metrics (where SEnt has midpack performance) are not suitable for assessing instance-wise coverage. In (b) we see that as we increase $\tau$ the proportion of lesions that are entirely undetected (a 'silent failure' of the UQ map) increases sharply, as detecting the smallest lesions comes at the cost of introducing a large number of TP and TN voxels as uncertain. The mean size of entirely undetected lesions is lowest among the methods that utilise Gaussian distributions over the logit space to model aleatoric uncertainty (Ind, SSN, SSN-Ens). Nonetheless all methods silently fail for between 30% to 50% of lesions at an equivalent UEO threshold (UEO = 0.4 in CVD dataset) in both in-distribution (CVD) and OOD (Challenge) data.

For the nnUnet variants, the SSN model does not improve over the SEnt model on patchwise scores C.1 (top) with slightly lower $p(acc|cert)$ across all thresholds. However SSN-Ens shows small improvements in $p(acc|cert)$ and PAvPU once adjusted for UEO score. While for the standard models SSN and SSN-Ens techniques show substantial improvement in the coverage and detection of small WMH over SEnt, for the nnUNet variants the improvements are limited but nonetheless consistent C.1 (bottom). However SSN-Ens and SSN show equivalent performance except for the size of undetected lesions once adjusted for UEO score.

While overall cohort statistics are useful, UQ performance varies greatly between individuals in UEO and coverage of missed lesions regardless of UQ method, making interpretation of specific thresholds difficult (See Appendix D). However, SSN-Ens attains the best Top AVD% score due to consistent performance across all volumes, while the performance of P-Unet degrades as WMH volume decreases (Fig. 6). The low sample diversity expressed in low volume subjects by P-Unet produces fewer uncertain but accurate voxels, yielding higher sUEO for low WMH volume subjects (c), while also explaining the high performance on the PAvPU metric and higher UEO at a given $\tau$ (Fig. 6).

### 4.2. UQ techniques robustly improve the prediction of the Fazekas score

In this section we examine the predictive power of spatial and uncertainty features extracted from the WMH probability and UQ outputs of the standard models in the Fazekas prediction task. Table 4 shows the results of the logistic regression classifier using features extracted from the outputs of different UQ methods.

Using only volume to predict Fazekas yields the worst performance across all metrics, with balanced accuracy of 0.62 and 0.71 for Deep and PV Fazekas respectively. Adding spatial features from the SEnt model always improves performance, with balanced accuracy of 0.67 and 0.73 in the Deep and PV regions respectively. Using spatial features from the WMH probability maps of P-UNet and SSN techniques alone does not always improve performance (e.g equivalent or worse Cohen's Kappa and balanced accuracy than using SEnt features for PVWMH). Adding uncertainty features from SSN-Ens always matches or improves performances over features from SEnt, with balanced accuracies of 0.74 for DWMH and 0.73 for PVWMH while always improving calibration (root brier score) with 0.65 for DWMH and 0.64 for PVWMH, compared to 0.72 and 0.66 for SEnt respectively. Including uncertainty features from P-UNet does not improve performance for DWMH, however improves performance across all metrics for PVWMH. The substantial performance improvements in the Deep WMH region when using uncertainty features from SSN-Ens may be explained by the difference in sample diversity and small WMH coverage in the UQ map between methods. While WMH in the PV region follow a common pattern of shape and location, small WMH in the Deep region are highly spatially varied, with the vast majority of small WMH in the Deep region and hence identification of possible WMH locations in DWMH implying a greater range of plausible DWMH volumes. This may be important when the implicit threshold between NAWM and WMH varies between annotators conducting a Fazekas annotation from MRI images.

The Fazekas classifiers show varying levels of sensitivity to the threshold $t$ and number of features $k$ hyperparameters selected. Supplementaries Appendix E and Appendix F show how balanced accuracy and root brier score varies across all hyperparameter settings tried across all methods. Across all thresholds the ordering of method performances is broadly consistent. Adding UQ features from the SSN Ens model always improves model calibration (root brier score) for both DWMH and PVWMH while reducing sensitivity to $k$ for the PVWMH balanced accuracy score. Adding UQ features improves balanced accuracy for both P-unet and SSN-Ens for PVWMH, albiet less consistently for P-Unet, which is particularly sensitive to the choice of $t$.

#### 4.2.1. Shift in annotator policy hinders assessment of generalisation of fazekas classifiers

Ideally, we should like to evaluate our models on another independent test set from a different cohort. We evaluated the performance of our models on the MSS3 dataset; for both PVWM and DWMH, performance appears to drop notably across all methods. However we observe that this is caused by difference in annotation policy followed by annotators across different datasets while generating the Fazekas scores. Fig. 7 demonstrates in-distribution test set vs MSS3 dataset confusion matrices for the SSN-Ens w/ UQ feature set and compares this to the WMH volume as a proportion of brain volume for each Fazekas category in both test sets. In the MSS3 dataset, the median

**Table 4**

Performance metrics for the Fazekas classifiers on the Deep and PV prediction tasks using features extracted from the outputs of each feature extraction method for the Cohen's Kappa, Balanced Accuracy, Area Under ROC and Root Brier Score metrics. **Volume Only:** Only the estimated WMH volume from the SEnt model is used as a feature. **SEnt/P-Unet/SSN-Ens:** Spatial and volumetric features are extracted from the WMH probability map output and used as input features to the Fazekas classifier. **(P-Unet/SSN Ens) w/ UQ** Spatial and volumetric features from the WMH probability and uncertainty map outputs are used as input features to the Fazekas classifier. The mean and 95% confidence intervals are calculated via bootstrapping using 1000 randomly sampled train/test splits.

| Target | Method | Kappa | Bal. Acc | AUROC | RBS |
| --- | --- | --- | --- | --- | --- |
| Deep WMH | 1. Volume Only | 0.37 (0.29, 0.45) | 0.62 (0.55, 0.70) | 0.83 (0.80, 0.86) | 0.74 (0.72, 0.77) |
|  | 2. SEnt | 0.44 (0.36, 0.53) | 0.67 (0.61, 0.74) | 0.86 (0.84, 0.89) | 0.72 (0.68, 0.75) |
|  | 3. P-Unet | 0.47 (0.38, 0.56) | 0.70 (0.62, 0.76) | 0.88 (0.85, 0.90) | 0.70 (0.66, 0.74) |
|  | 4. P-Unet w/ UQ | 0.47 (0.38, 0.57) | 0.69 (0.61, 0.76) | 0.88 (0.85, 0.91) | 0.70 (0.66, 0.74) |
|  | 5. SSN Ens | 0.50 (0.42, 0.58) | 0.71 (0.63, 0.77) | 0.89 (0.86, 0.91) | 0.68 (0.65, 0.72) |
|  | 6. SSN Ens w/ UQ | **0.57** (0.48, 0.66) | **0.74** (0.66, 0.81) | **0.91** (0.88, 0.93) | **0.65** (0.60, 0.70) |
| PV WMH | 1. Volume Only | 0.51 (0.42, 0.59) | 0.71 (0.64, 0.78) | 0.89 (0.87, 0.92) | 0.68 (0.65, 0.71) |
|  | 2. SEnt | 0.54 (0.46, 0.63) | 0.73 (0.66, 0.79) | 0.90 (0.88, 0.93) | 0.66 (0.62, 0.70) |
|  | 3. P-Unet | 0.54 (0.46, 0.63) | 0.73 (0.66, 0.80) | 0.90 (0.88, 0.93) | 0.66 (0.62, 0.70) |
|  | 4. P-Unet w/ UQ | **0.58** (0.48, 0.67) | **0.74** (0.67, 0.81) | **0.91** (0.89, 0.94) | **0.64** (0.60, 0.69) |
|  | 5. SSN Ens | 0.55 (0.45, 0.64) | 0.72 (0.65, 0.79) | **0.91** (0.88, 0.93) | 0.65 (0.61, 0.70) |
|  | 6. SSN Ens w/ UQ | 0.56 (0.48, 0.65) | 0.73 (0.66, 0.79) | **0.91** (0.89, 0.94) | **0.64** (0.60, 0.69) |

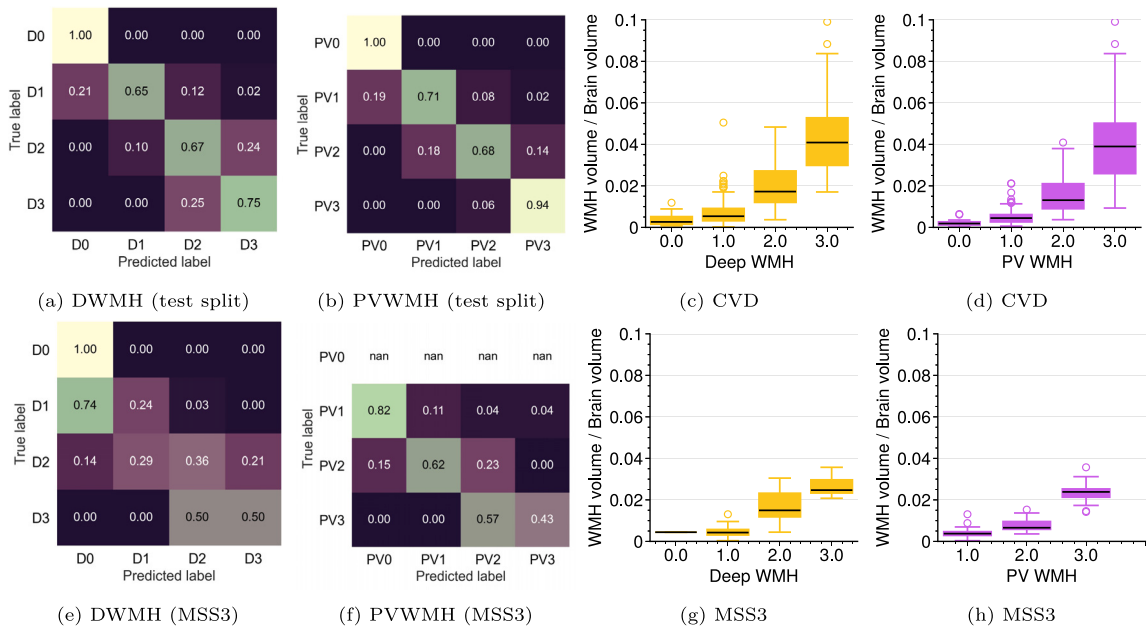

**Fig. 7.** Comparison of in distribution and out of distribution Fazekas classification with the distribution of WMH volumes per fazekas class. (a, b, e, f): Confusion matrix of the Fazekas classifier (from a single random train/test sample) using spatial and volumetric features from the WMH probability and uncertainty maps of the SSN-Ens model (SSN-Ens w/ UQ) on in distribution ((a) DWMH, (b) PVWMH) and out of distribution ((MSS3) (e) DWMH, (f) PVWMH) data. (c) and (d): Distribution of WMH volume as a proportion of brain volume per class for the CVD dataset. (g) and (h): Distribution of WMH as a proportion of brain volume for the MSS3 dataset.

WMH proportion for DWMH 0 class subjects is approximately equal to that of the DWMH 1 median WMH proportion, while the distribution of volumes of all classes > 0 are shifted lower compared to the CVD dataset. Consequently, most DWMH 1 images are as Fazekas 0 while confusion between each true class $c > 0$ and class $c - 1$ is increased. For PVWMH, the difference in WMH volume per class $c$ in MSS3 is so severe that each class's volume distribution is approximately equivalent to the distribution of class $c - 1$ in the CVD dataset, with no class 0 present in the dataset. Hence, the model predicts a Fazekas score 1 class lower than the ground truth in the vast majority of cases and we conclude that the model does generalise well to the MSS3 dataset given the rater policy disagreement.

*4.3. UQ techniques improve quality control and volume estimation of WMH segmentation*

We now examine the QC classification task. Table 5 demonstrates the performance metrics for each method. Similarly to the Fazekas score, using volume information is the least informative feature set for detecting poor quality segmentations with a balanced accuracy of 0.74 and root brier score of 0.60. Crucially, using UQ features from either the SSN-Ens model or P-Unet model substantially improve the kappa, balanced accuracy and root brier score, with the inclusion of UQ features from SSN-Ens improving balanced accuracy to 0.82 and root brier score to 0.50. The inclusion of UQ features from either P-Unet or SSN-Ens improves root brier score regardless of hyperparameters $t$ and $k$, however for all methods balanced accuracy varies with $t$ (Appendix G). Fig. 8 explores how the distribution of WMH volumes predicted by the stochastic methods varies as the ground truth WMH volume changes. At low WMH volumes all methods consistently overestimate the WMH burden in the median segmentation. However, methods with low $D^2_{\text{GED}}$ exhibit high sample diversity, such that the distribution of volumes predicted by the model is more likely to include the ground truth. P-Unet exhibits minimal sample diversity at individuals with low WMH volume, while Ind exhibits almost none across the whole range of volumes, as expected. Mc-Dropout shows a broader range, but like P-Unet and Ind, if the median is far from zero, the range of possible volumes predicted cannot match the ground truth. Furthermore, the skewness of the distribution of WMH volumes from SSN-Ens increases as volumes are smaller (Fig. 6). Hence we can correct for poor model performance at low volumes automatically by selecting a low volume segmentation from the predicted WMH distribution. This explains why the UQ representation from SSN-Ens can be used to effectively detect poor quality segmentation, while P-Unet UQ features with low sample diversity do not yield equivalent gains.

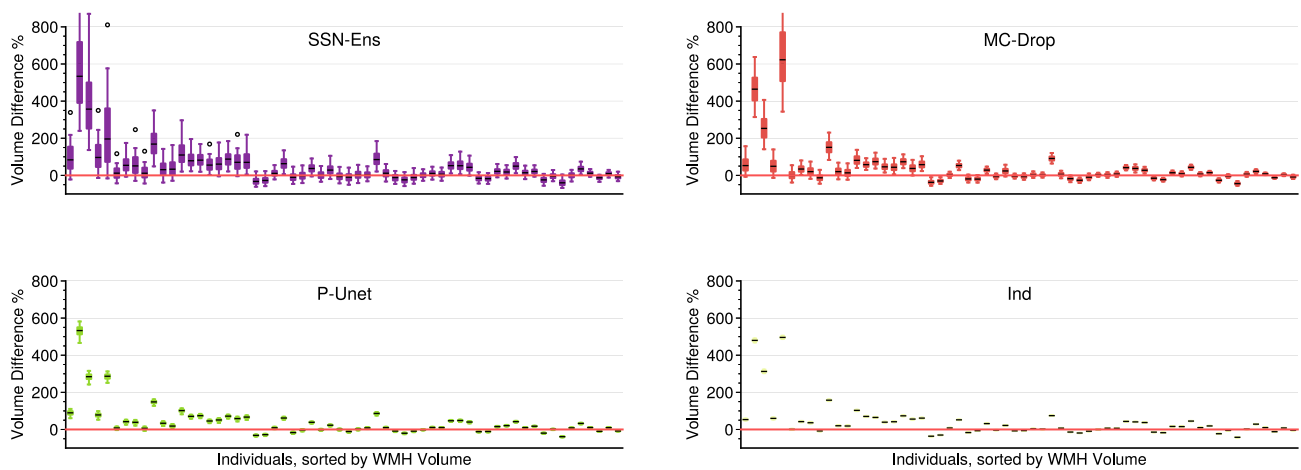

**Fig. 8.** Volume Difference distribution of samples produced by (stochastic) methods on the OOD (Challenge) dataset. Individuals in the dataset are sorted in ascending order of ground truth WMH volume from left to right. Red Line indicates volume difference of 0 (i.e ground truth WMH volume).

**Table 5**
Performance metrics for the QC classifiers using features extracted from the outputs of each feature extraction method for the Cohen's Kappa, Balanced Accuracy, Area Under ROC and Root Brier Score metrics. **Volume Only:** Only the estimated volume from the SEnt model is used as a feature. **SEnt/P-Unet/SSN-Ens:** Spatial and volumetric features are extracted from the WMH probability map outputs and used as input features to the QC classifier. **(P-Unet/SSN Ens) w/ UQ** Spatial and volumetric features from the WMH probability and uncertainty map outputs are used as input features to the QC classifier. The mean and 95% confidence intervals are calculated via bootstrapping using 1000 randomly sampled train/test splits.

| Method | Kappa | Bal. Acc | AUROC | RBS |
|---|---|---|---|---|
| 1. Volume Only | 0.34 (0.21, 0.47) | 0.74 (0.66, 0.82) | 0.86 (0.77, 0.94) | 0.60 (0.54, 0.67) |
| 2. SEnt | 0.45 (0.30, 0.60) | 0.79 (0.69, 0.87) | 0.88 (0.80, 0.96) | 0.54 (0.48, 0.61) |
| 3. P-Unet | 0.40 (0.25, 0.56) | 0.77 (0.67, 0.85) | 0.88 (0.79, 0.95) | 0.56 (0.49, 0.63) |
| 4. P-Unet w/ UQ | 0.54 (0.38, 0.70) | **0.82 (0.72, 0.91)** | 0.88 (0.79, 0.95) | 0.52 (0.43, 0.60) |
| 5. SSN Ens | 0.44 (0.30, 0.58) | 0.79 (0.70, 0.87) | 0.88 (0.79, 0.95) | 0.56 (0.49, 0.63) |
| 6. SSN Ens w/ UQ | **0.55 (0.38, 0.73)** | **0.82 (0.71, 0.91)** | **0.89 (0.80, 0.96)** | **0.50 (0.42, 0.58)** |

*4.4. SSN-Ens provides semantically meaningful UQ maps that identify stroke lesions and segmentation errors*

So far we have shown that UQ techniques improve the robustness and generalisation of WMH segmentation models, Fazekas classification and identification of poor quality images. We now examine the semantic information captured by the UQ maps of the top performing method (SSN-Ens) vs the baseline SEnt in the MSS3, Challenge and ADNI datasets in first the standard and then nnUNet model variants.

We first examine images with known stroke lesions in the MSS3 dataset (Fig. 9(a) and Appendix I). SSN-Ens is able to highlight areas where there is a stroke lesion within the spatial distribution of WMH as uncertain. Crucially, SSN-Ens highlights stroke lesions (that lie within the spatial and shape distribution of WMH) as uncertain. However SEnt is likely to segment stroke lesions as WMH and often erroneously identifies only the boundary of the lesion as uncertain. SEnt may even erroneously highlight the cavitated region of stroke lesions as uncertain Appendix I.2b, when these are clearly not WMH, leaving the rest of the lesion certain. While this suggests uncertainty quantification may be able to improve discerning between WMH and stroke lesions, stroke lesions outside the spatial and shape distribution of WMH are, as expected, not segmented and not uncertain (I.2a, I.3b), and hence require training models with additional stroke lesion labels. Due to differences in acquisition parameters in the MSS3 compared with the images from the CVD dataset, in the MSS3 gray matter appears brighter relative to the white matter in the FLAIR images. This causes confusion for all models, with gray matter around the giri, insular cortex and caudate nucleus occasionally uncertain, and segmented as WMH in rare cases. However we note that SSN-Ens is substantially less prone to this behaviour, almost never segmenting gray matter and reporting less gray matter as highly uncertain, showing improved robustness to real world OOD scenarios (I.1a, I.1c).

However, when we compare the nnUNet variants for both SSN-Ens and SEnt, the confusion between subcortical stroke lesions and WMH is increased (Appendix J.1). Both models often partially or entirely segmenting stroke lesions as WMH while remaining confident in the prediction, reducing the clinical and research utility of the models. Nonetheless, the SSN-Ens variant is more likely to avoid segmenting stroke lesions, show higher uncertainty within stroke lesions and higher confidence when WMH are correctly segmented. Both nnUnet models highlight gray matter and normal appearing white matter as uncertain in the images, with the SEnt variant on occasion segmenting these regions, while the SSN-Ens model is noticeably less uncertain in these regions. However both models highlight considerably less gray and normal appearing white matter as uncertain.

Next we examine the identification of small WMH in the Challenge dataset (Fig. 9(b) and Appendix K). While both models (i.e., SEnt and SSN-Ens) may fail to segment small WMH, SSN-Ens may identify whole clusters of unsegmented WMH in the uncertainty map K.1b. Furthermore, SSN-Ens is more likely to highlight false positives in the segmentation as uncertain (K.1a, K.2b), while remaining confident in the centre of correctly segmented lesions. SEnt may however be highly uncertain among correctly segmented lesions (i.e SSN-Ens is not simply always more uncertain about a given region than SEnt). However, we find SSN-Ens reports low levels of uncertainty for some areas of gray matter where SEnt does not; this behaviour is specific to the differences

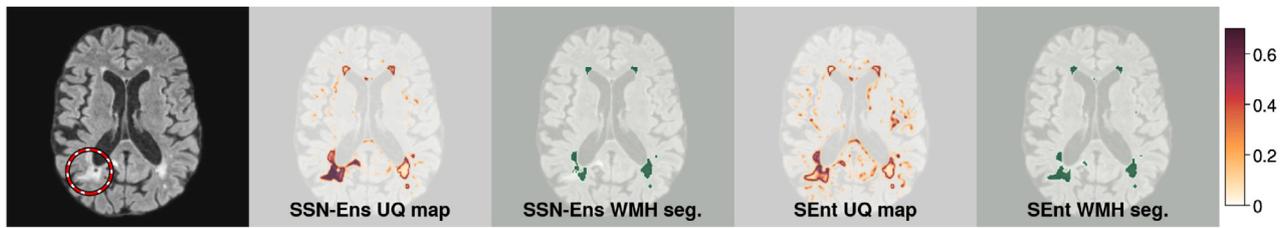

(a) **SEnt fails to identify a stroke lesion (segmented as WMH) as uncertain.** Red ring denotes known stroke lesion. While SSN-Ens segments part of the stroke lesion erroneously, the entire stroke lesion is identified as (highly) uncertain. SEnt only identifies the lesion boundary as uncertain, even when the stroke lesion deviates outside the spatial distribution of WMH. SEnt also highlights areas of the insular cortex as uncertain and partially segments this region, while SSN-Ens greatly reduces the uncertain area and does not segment the insular cortex.

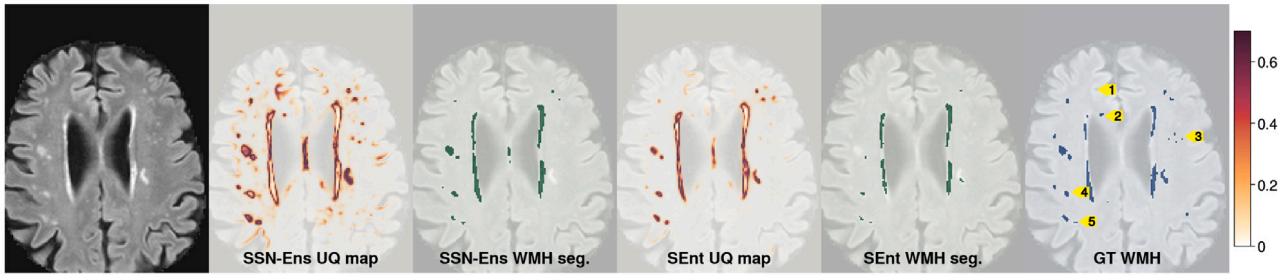

(b) **Arrow 1**: SSN-Ens introduces a number of areas of gray matter as uncertain in the image. **Arrow 2** and **Arrow 3**: Both models fail to segment a number of small WMH clusters. However, SSN-Ens more consistently identifies these WMH as uncertain. **Arrow 4**: Both methods segment this WMH. However, the SEnt model is uncertain about the entire lesion. While this is not a failing of the uncertainty map, an ideal model will be certain in this case. The SSN-Ens model identifies only the boundary of this WMH as uncertain, and is confident in its prediction for the centre of the WMH. **Arrow 5**: The same as **Arrow 4**, for a cluster of two WMH. SSN-Ens highlights a number of gray matter areas as uncertain in this image.

Fig. 9. Qualitative Analysis: Comparison of the WMH probability map thresholded at 0.5 (WMH seg.) and UQ maps of SSN-Ens and Sent, using examples from the (a) MSS3 and (b) Challenge datasets. Colourbars indicate UQ map values.

in acquisition protocols and requires filtering out low level uncertainty values K.2d. Nonetheless, both models can silently fail (neither segment nor consider uncertain) some WMH.

The nnUnet variants show less differences in the uncertainty maps than for the standard models, however semantic differences remain (Appendix L.1). The SEnt nnUNet model is more often highlights gray matter and bright normal appearing white matter as uncertain and may segment it in rare cases. While the SSN-Ens model is typically still somewhat uncertain in these areas, the model is considerably more confident (correctly) than the SEnt model and crucially is more confident than within unsegmented WMH than around gray and normal appearing white matter. The opposite can be true for the SEnt model, making thresholding of the uncertainty map difficult. The SSN-Ens nnUNet model is also more likely to detect in the uncertainty map WMH missed in the segmentation.

The standard and nnUNet SSN-Ens models are more likely to highlight areas of white matter with elevated intensity. These regions may be borderline WMH or could develop into WMH over time, however their presence in the uncertainty map requires a higher uncertainty threshold to achieve the same uncertainty error overlap score, due to these regions not being present in the binary segmentation, thus complicating the threshold and metric based analysis of the uncertainty map quality produced by different techniques.

Finally, examination of images with artefacts unseen during training from the ADNI dataset reveals model failures. Introduction of a bias field to the FLAIR image (Fig. 10(a)) causes both methods to produce high levels of uncertainty around bright regions of the image while failing to detect WMH in darker regions of the image. The segmentations of both methods are highly asymmetrical compared to the WMH visible in FLAIR and SEnt introduces erroneous WMH on the boundary of the fourth ventricle. We further find that in all cases where a visible head motion artefact is present (Fig. 10(b)), UQ maps highlight the bright vertical streaks that overlap with the intensity distribution of WMH as uncertain. While the uncertainty map highlights the error, clearly a model that is robust to head motion and bias field is preferable, and modelling uncertainty alone is insufficient to achieve such robustness.

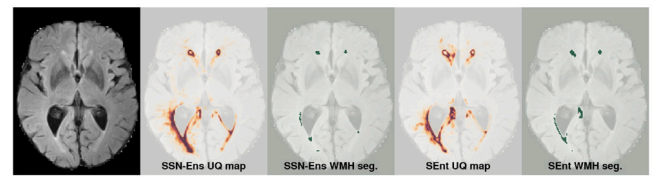

(a) Bias field present in FLAIR. Both methods fail to account for the bias field in this image, oversegmenting the right hemisphere. While SSN-Ens highlights large areas of the right occipital caps as uncertain, SEnt obtains a poorer segmentation, erroneously segmenting more white matter with greater asymmetry between hemispheres. SEnt incorrectly identifies a white matter region on the boundary of the fourth ventricle as WMH - only some of this segmented region is uncertain.

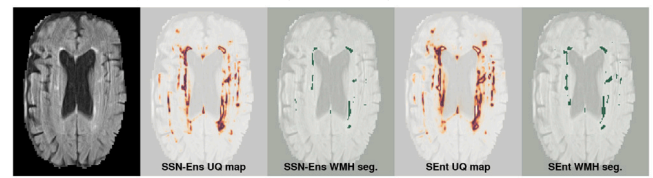

(b) Severe head motion artefact. Both methods highlight artefacts due to head motion as uncertain and incorrectly segment some of the affected regions. Both methods segment plausible WMH, however SEnt incorrectly (and sometimes confidently) segments an increased number of motion artefacts.

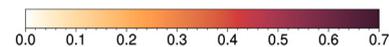

(c) Colourbar of uncertainty value (entropy) in UQ maps (ranging from 0 to ln 2)

Fig. 10. Comparison of the WMH probability map thresholded at 0.5 (WMH seg.) and UQ maps of SSN-Ens and SEnt models, demonstrating failures on OOD data in the ADNI dataset. (a) Bias field in FLAIR. (b) Severe head motion with bright streaks in FLAIR.

The nnUNet variants also perform poorly in the bias field and head motion examples (Appendix M.1), with both segmenting normal white matter in the bias field case and showing bright regions of the head motion artefact in the uncertainty map. The SSN-Ens model reduces the volume of normal white matter segmented over the SEnt model, while the SEnt is certain even where the segmentation is likely incorrect. However the SSN-Ens nnUNet variant highlights noticeably more of the bright regions of the head motion artefact as uncertain compared to the SEnt model.

## 5. Discussion

In this work we present four main contributions:

**(1)** We benchmark uncertainty modelling techniques for semantic segmentation, finding that UQ improves WMH segmentation performance in both in-distribution and OOD data, and identifies areas where small WMH may have been missed. The combination of ensembles and stochastic segmentation networks (SSN-Ens) gives the strongest results, generalising robustly to unseen data.

**(2)** We propose an automated classification method for the Fazekas score using spatial and volumetric features from WMH probability and UQ maps. Spatial features significantly improve Fazekas classification performance over simple volume based estimation, and leveraging uncertainty information from SSN-Ens consistently improves classification performance for in-distribution data

**(3)** We demonstrate that features from SSN-Ens UQ maps enhance poor quality image detection, and that UQ techniques that yield high sample diversity can allow automation correction when WMH volume is overestimated in images from subjects with low WMH volume.

**(4)** We qualitatively examine the semantic information content of uncertainty maps from SSN-Ens compared to a baseline deterministic model (SEnt). SSN-Ens identifies stroke lesions within the spatial and volume distribution of WMH and identifies false negative WMH clusters as uncertain, while reducing the impact of some artefacts and acquisition parameter changes. This greatly improves over the baseline SEnt, which often misses entire WMH clusters in the deep white matter and can produce nonsensical or irrelevant uncertainty information.

### 5.1. Assessment of UQ techniques

The implications of our findings are multifold. When comparing UQ techniques, it is important to consider multiple metrics, and per-subject performance. Strong performance on a single metric may mask other semantic properties of the uncertainty map (e.g SSN-Ens attains higher scores across all thresholds than P-Unet for patchwise metrics such as $p(acc|cert)$, however attains noticeably lower UEO at the same thresholds) (Fig. 5). Metrics such as AVD and sUEO vary considerably across subjects (Fig. 6, Appendix D), so per-subject evaluation is key when choosing the right method for downstream tasks. For example, P-Unet will be preferable in a human-in-the-loop annotation setting where an annotator corrects uncertain voxels in the image, due to its high UEO and PAvPU scores. For accurate volume estimation and identification of all potential WMH lesions, SSN-Ens will be preferable due to its high sample diversity and hence consistent coverage of the true WMH volume in the predicted distribution of WMH volumes.

Given the high aleatoric uncertainty inherent to WMH segmentation, and hence the high sample diversity required to capture the 'ground truth' segmentation within the predicted WMH distribution, $D^2_{\text{GED}}$ is dominated by the sample diversity term, which reduces the sUEO score. We found P-Unet is insufficiently expressive to capture the smallest WMH lesions while unable to consistently produce segmentations close to the WMH volume (Fig. 8), and adjusting the latent space size did not resolve this. Further techniques could improve P-Unet's expressivity, for example: using normalising flows (Valiuddin et al., 2021), hierarchical latent spaces (Kohl et al., 2019) and using multivariate Gaussian mixtures (Bhat et al., 2022) or incorporating epistemic uncertainty (Hu et al., 2019). However, such approaches increase the complexity of the method considerably compared to the SSN, which outperforms P-Unet with better quality in the distribution of WMH segmentations produced (e.g., lower $D^2_{\text{GED}}$, higher Top Dice score).

Implementation of explicit uncertainty quantification techniques improves the quality of uncertainty map outputs regardless of the specific choices of network architecture and training paradigm (Table 3), nonetheless we found that when using nnUNet as opposed to a simpler UNet setup the segmentation performance gap between the best UQ technique and the baseline is small Table 2 However the choice of architecture does impact the semantic properties of the model, with nnUNet variants less effective at discriminating stroke lesions from WMH Appendix J while increasing the overestimation of the WMH volume.

Another popular approach for modelling aleatoric uncertainty is test time augmentation (TTA) (Wang et al., 2019), where multiple samples are generated by applying augmentations to an image before applying the model. While interpolations during augmentation may affect the delineation of the WMH boundaries, we conclude that intuitively the ambiguity around the presence of a small WMH (for example) should not depend on the typical augmentations used (such as rotations, rescaling and reflections). Hence we avoid comparing this type of methods in this work.

The assessment of the robustness of UQ techniques should be performed on realistic OOD data. Various works show that evidential deep learning may improve robustness (Amini et al., 2020; Zou et al., 2022, 2023b) while remaining computationally efficient. A common approach to demonstrate such robustness is via perturbations such as noise and augmentations, with similar approaches for MC-Drop (Mojiri Forooshani et al., 2022). However, when using a separate real world OOD data, we find that performance improvements over the baseline from evidential deep learning (Evid) are limited compared to both MC-Drop and ensembling techniques, which achieve lower AVD (Table 2) and lesion coverage (Fig. 5). Hence using OOD data to guide model choice is essential, as augmentations may not capture acquisition protocol, demographic and image quality variability. Furthermore, where robustness in terms of domain generalisation is considered the principal benefit of UQ, other UQ methods should not be the ultimate yardstick for method comparison, where arguably more appropriate baselines in the literature regarding robustness and domain generalisation are available (Lee et al., 2018; Zhao et al., 2021).

### 5.2. Relevance of UQ for generating Fazekas scores and quality control

UQ techniques capture semantically meaningful spatial information for informing the Fazekas classification task (Table 4). Our proposed approach leveraging spatial and volumetric features across different brain regions clearly improves Fazekas scores' prediction over global WMH volume alone as it aligns naturally with neuroradiological interpretation guidelines. However, since Fazekas is generated directly from inspecting MRI images, factors such as the relative intensity of the normal appearing white matter (NAWM) in the WMH penumbra, marginally hyperintense regions or ambiguous small deep isolated WMH may contribute to differences in the SVD severity perceived by an annotator and hence the Fazekas score. Uncertainty maps help capture such regions, and uncertainty in the WMH penumbra that is anterior or posterior to the ventricle horn may give improved detection of potential confluence between Deep and PV regions. This may explain why UQ map features improve both model accuracy and calibration (lower RMBS score) in in-distribution data, as the model can express uncertainty among cases that are in-between Fazekas categories. However, we have demonstrated that evaluating the generalisability of Fazekas score classifiers in OOD data is potentially ill-posed due to the different criteria perceived in the annotation of each datasets and the lack of standards for generating consistent ground truth annotations that are amenable to probabilistic modelling. Where annotators use different criteria (policies) when providing annotations, this hinders cross-dataset comparison and can cause misleading results (Philps et al., 2024). For example, in addition to the differences in volume distribution of each score between studies that we have demonstrated, a single small WMH cluster in the deep white matter may be considered by one annotator as sufficient for giving Fazekas DWMH score of 1 whilst another annotator might have chosen to give the score of 0. Hence, despite the known effectiveness of machine learning approaches, uniformity in the criteria for generating Fazekas scores, for example, based on a WMH segmentation agreed by annotators as gold standard, could allow for a clearer and interpretable link between the uncertainty in WMH and the downstream uncertainty in the Fazekas score. Despite the known floor and ceiling effects of the clinical visual scoring systems (Olsson et al., 2013; van Straaten et al., 2006-03), effective automated Fazekas estimation could facilitate larger-scale research into WMH in clinical data.

Despite segmentation quality (e.g Dice score) depending heavily on WMH volume, we expect spatial features to improve QC prediction performance due to the increased difficulty in Deep WMH segmentation compared to the PV region. We would also expect that high uncertainty in a region may be indicative of low segmentation quality, with UQ features from both P-Unet and SSN-Ens improving error detection (Table 5). However, using UQ features from P-Unet was less effective for detecting low quality images than those from SSN-Ens. This may suggest that among subjects where the median predicted WMH volume is greater than the ground truth, the high sample diversity and positive skew of the predicted WMH distribution more evident from SSN-Ens predictions may improve identification of poor quality images and help predict the DWMH Fazekas score.

### 5.3. Future avenues for research

There are many avenues for future research. Longitudinal analysis of uncertain regions in the NAWM could reveal meaningful biological information, such as regions of the normal appearing white matter that later develop into WMH, particularly around small isolated WMH.

Disentangling epistemic uncertainty from aleatoric uncertainty may allow for uncertainty information to be more effectively targeted at quality control and related tasks such as OOD detection which intuitively depend on epistemic uncertainty. However, disentangling epistemic from aleatoric uncertainty is a non-trivial task, as the conventional decomposition (Houlsby et al., 2011) of Shannon entropy (total uncertainty) into mutual information (epistemic uncertainty) and conditional entropy (aleatoric uncertainty) can yield inconsistent results in practice (Wimmer et al., 2023). Hence, in this work, we have chosen to evaluate the utility of the total uncertainty expressed by each technique without disentangling the epistemic component. However aleatoric uncertainty may not be informative for quality control tasks, while epistemic uncertainty due to artefacts, such as head motion, will impact the utility of UQ maps for other downstream tasks such as Fazekas prediction. This indicates that further separation of epistemic and aleatoric uncertainty is warranted.

Given the semantic differences between nnUNet and the standard model variants, the impact of training paradigm should be further explored, in particular how the choices regarding data augmentation during training may effect the utility of the uncertainty maps produced. Furthermore, given the limited availability of heterogenous training data for WMH segmentation, augmentation strategies specific to MRI should be explored to bridge the performance gap in images with different acquisition parameters or imaging artefacts.

The SSN-Ens method could be further improved in a number of ways. First, modelling the uncertainty over the logit space with other long tailed distributions, such as a low rank multivariate $T$ distribution could further reduce the impact of noise during training (Gonzalez-Jimenez et al., 2023). Secondly, one hypothesis for the performance improvements on OOD (out-of-distribution) data attained with ensembles is due to ensemble diversity (de Mathelin et al., 2023). Hence, methods that attempt to improve the diversity of the ensemble have been pursued, demonstrating improvements in robustness and OOD detection (Mehrtens et al., 2022; Zaidi et al., 2021; D'Angelo et al., 2021). However, some argue this is explained due to differences in model capacity (Ovadia et al., 2019). Hence, further work comparing the effectiveness of SSN-Ens to other techniques that improve robustness to OOD data should be explored. Nonetheless, improving the diversity of the ensemble may be an effective way to boost robustness and uncertainty representations. One way to do this, specific to WMH segmentation, may be to train individual ensemble elements on different datasets that cover a range of WMH segmentation policies (Philps et al., 2024), so that the full range of annotator disagreement (and thereby aleatoric uncertainty) can be accurately reflected by the model. However, complete sets of desired MRI acquisitions for applying a model (e.g FLAIR and T1w) are in general not available in clinical settings (Ruffle et al., 2023) and so assessing suitability for clinical settings needs to go beyond assessing OOD performance, for example by providing segmentation on available scans, whether it is T2, T2* or T1-CE for broad applicability. Finally, in order to move towards an automated neuroradiological support system for studying SVD in clinical patients, application of UQ and robust generalisation techniques to other neuroradiological markers of SVD in a multi-label setting, in particular for stroke lesions and perivascular spaces, is essential. This would not only improve stroke lesion detection but allow separation of uncertainty around WMH locations and size from uncertainty around pathology type, improving performance in downstream tasks such as Fazekas scoring.

**CRediT authorship contribution statement**

**Ben Philps:** Writing – original draft, Software, Methodology, Investigation, Formal analysis, Data curation, Conceptualization. **Maria del C. Valdés Hernández:** Writing – review & editing, Writing – original draft, Supervision, Resources, Methodology, Data curation, Conceptualization. **Chen Qin:** Writing – review & editing, Supervision, Conceptualization. **Una Clancy:** Writing – review & editing, Data curation. **Eleni Sakka:** Writing – review & editing, Data curation. **Susana Muñoz Maniega:** Writing – review & editing, Data curation. **Mark E. Bastin:** Writing – review & editing, Data curation. **Angela C.C. Jochems:** Writing – review & editing, Data curation. **Joanna M. Wardlaw:** Writing – review & editing, Data curation. **Miguel O. Bernabeu:** Writing – review & editing, Supervision, Methodology, Conceptualization.

**Declaration of competing interest**

The authors declare that they have no known competing financial interests or personal relationships that could have appeared to influence the work reported in this paper.

**Acknowledgements**

We thank Simon R. Cox, Daniela Jaime Garcia, Junfang Zhang, Xiaodi Liu and Yajun Cheng for their contribution to the curation of datasets utilised in this work. BP was funded by the United Kingdom Research and Innovation Centre for Doctoral Training in Biomedical AI Programme scholarships (grant EP/S02431X/1). For the purpose of open access, the author has applied a creative commons attribution (CC BY) licence to any author accepted manuscript version arising. Funding from Row Fogo Charitable Trust, United Kingdom (Ref No: AD.ROW4.35. BRO-D.FID3668413), and the UK Medical Research Council (UK Dementia Research Institute at the University of Edinburgh, award number UK DRI-4002;G0700704/84698) are also gratefully acknowledged. M.O.B. gratefully acknowledges funding from:

Fondation Leducq Transatlantic Network of Excellence (17 CVD 03); EPSRC, United Kingdom grant no. EP/X025705/1; British Heart Foundation and The Alan Turing Institute Cardiovascular Data Science Award (C-10180357); Diabetes UK, United Kingdom (20/0006221); Fight for Sight, United Kingdom (5137/5138); the SCONe projects funded by Chief Scientist Office, Edinburgh & Lothians Health Foundation, Sight Scotland, the Royal College of Surgeons of Edinburgh, United Kingdom, the RS Macdonald Charitable Trust, and Fight For Sight; the Neurii initiative which is a partnership among Eisai Co., Ltd, Gates Ventures, LifeArc and HDR UK.

Data collection and sharing for the Alzheimer's Disease Neuroimaging Initiative (ADNI) is funded by the National Institute on Aging (National Institutes of Health Grant U19 AG024904). The grantee organization is the Northern California Institute for Research and Education. In the past, ADNI has also received funding from the National Institute of Biomedical Imaging and Bioengineering, United States, the Canadian Institutes of Health Research, and private sector contributions through the Foundation for the National Institutes of Health (FNIH) including generous contributions from the following: AbbVie, Alzheimer's Association; Alzheimer's Drug Discovery Foundation, United States; Araclon Biotech; BioClinica, Inc.; Biogen; Bristol-Myers Squibb Company; CereSpir, Inc.; Cogstate; Eisai Inc.; Elan Pharmaceuticals, Inc.; Eli Lilly and Company, United States; EuroImmun; F. Hoffmann-La Roche Ltd and its affiliated company Genentech, Inc.; Fujirebio; GE Healthcare, United Kingdom; IXICO Ltd.; Janssen Alzheimer Immunotherapy Research & Development, LLC.; Johnson & Johnson Pharmaceutical Research &Development LLC.; Lumosity; Lundbeck, Denmark; Merck & Co., Inc., United States; Meso Scale Diagnostics, LLC.; NeuroRx Research; Neurotrack Technologies; Novartis Pharmaceuticals Corporation, United States; Pfizer Inc.; Piramal Imaging; Servier, France; Takeda Pharmaceutical Company, Japan; and Transition Therapeutics.

**Appendix A. Supplementary data**

Supplementary material related to this article can be found online at https://doi.org/10.1016/j.media.2025.103697.

**Data availability**

The source-code and links to models can be found on GitHub (see manuscript). The datasets referred to as 'CVD' and 'MSS3' can be requested to the primary studies' data holders.

# Appendix A. Flair Intensity pre and post normalization

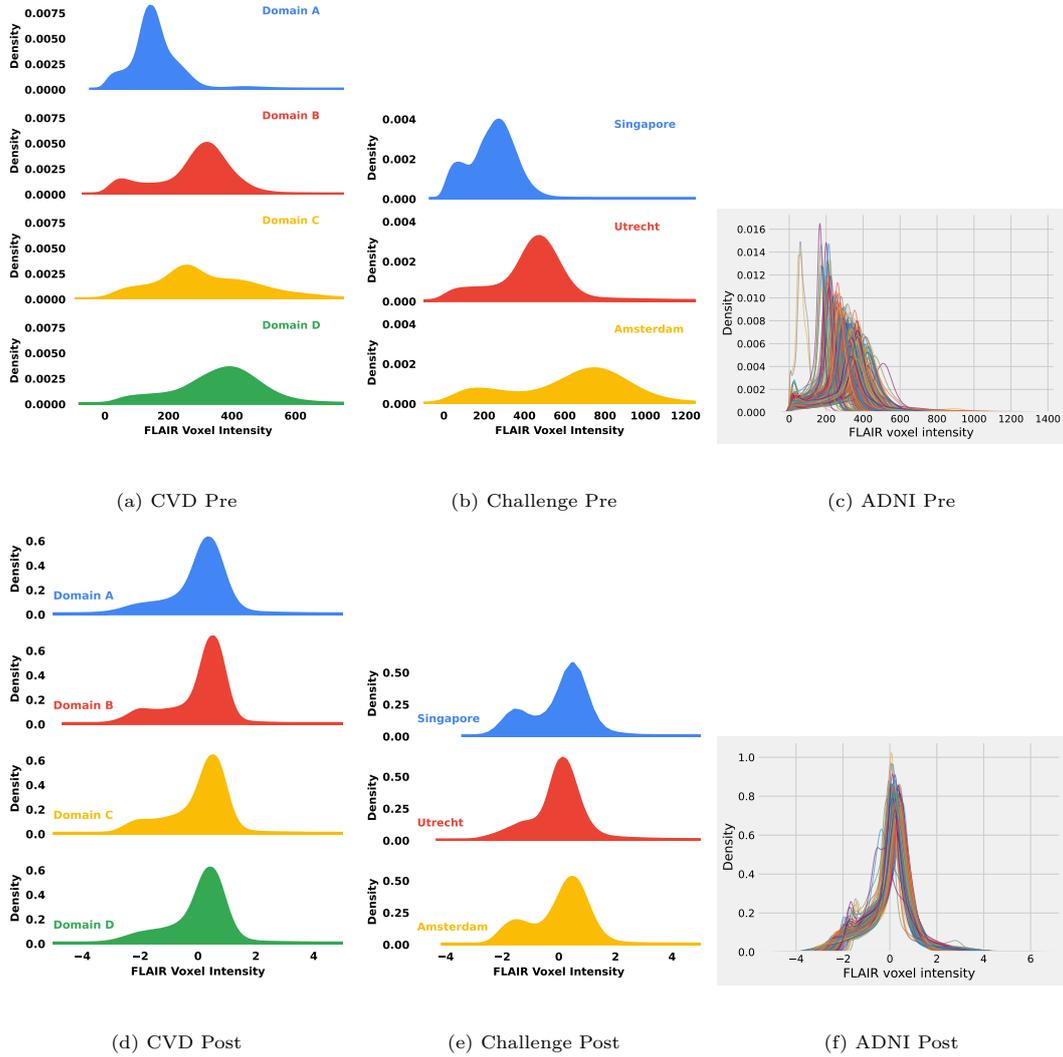

Figure (A.1) FLAIR voxel intensity distributions for all datasets pre and post Z-score normalization. For CVD and WMH Challenge datasets, we show a separate distribution for each domain, for ADNI we plot a distribution per image.

# Appendix B. MRI Acquisition Parameters

Table (B.1) MRI parameters for each dataset.

| Dataset | Institute | Scanner | T1-weighted Voxel size (mm3) | TE/TR/TI (ms) | FLAIR Voxel size (mm3) | TE/TR/TI (ms) | Total |
|---|---|---|---|---|---|---|---|
| CVD | CCBS | 1.5T GE Signa HDxt | 0.9 × 0.9 × 6.5 | 9.0/ 440/ - | 0.9 × 0.9 × 6.5 | 147/9002/ 2200 | 51 |
|  |  |  | 0.5 × 0.5 × 4.0 | 4.0/ 9.6/ 500 | 0.5 × 0.5 × 4.0 | 100/8000/ 2000 | 25 |
|  |  |  | 1.0 × 1.3 × 1.0 | 4.0/ 9.7/ 500 | 1.0 × 1.0 × 3.0 | 140/9000/ 2200 | 62 |
|  |  |  | 1.0 × 0.9 × 1.0 | 4.0/ 9.7/ 500 | 0.5 × 0.5 × 6.0 | 140/9000/ 2200 | 112 |
| MSS3 | CCBS | 3T Siemens Prisma | 1.0 × 1.0 × 1.0 | 4.4/ 2500/ - | 1.0 × 1.0 × 1.0 | 388/5000/ 1100 | 65 |
| Chal. | UTC | 3T Philips Achieva | 1.0 × 1.0 × 1.0 | 4.5/ 7.9/ - | 1.0 × 1.0 × 3.0 | 125/11000/ 2800 | 20 |
|  | SING | 3T Siemens TrioTim | 1.0 × 1.0 × 1.0 | 1.9/ 2300/ 900 | 1.0 × 1.0 × 3.0 | 82/9000/ 2500 | 20 |
|  | AMS | 3T GE Signa HDxt | 0.9 × 0.9 × 1.0 | 3.0/ 7.8/ - | 1.0 × 1.0 × 1.2 | 126/8000/ 2340 | 20 |
| ADNI | D1 | 3T Philips Intera | 1.0 × 1.0 × 1.2 | 3.2/ 6.8/ - | 0.9 × 0.9 × 5.0 | 90/9000/ 2500 | 16 |
|  | D2 | 3T Philips Intera | 1.0 × 1.0 × 1.2 | 3.1/ 6.8/ - | 0.9 × 0.9 × 5.0 | 90/9000/ 2500 | 3 |
|  | D3 | 3T Philips Ingenia | 1.0 × 1.0 × 1.2 | 3.2/ 6.8/ - | 0.9 × 0.9 × 5.0 | 90/9000/ 2500 | 9 |
|  | D4 | 3T Philips Achieva | 1.0 × 1.0 × 1.2 | 3.1/ 6.8/ - | 0.9 × 0.9 × 5.0 | 90/9000/ 2500 | 34 |
|  | D5 | 3T Philips Achieva | 1.0 × 1.0 × 1.2 | 3.2/ 6.8/ - | 0.9 × 0.9 × 5.0 | 90/9000/ 2500 | 6 |
|  | D6 | 3T Philips Achieva | 1.0 × 1.0 × 1.2 | 3.1/ 6.7/ - | 0.9 × 0.9 × 5.0 | 90/9000/ 2500 | 10 |
|  | D7 | 3T Siemens Biograph mMR | 1.0 × 1.0 × 1.2 | 3.0/ 2300/ 900 | 0.9 × 0.9 × 5.0 | 91/9000/ 2500 | 6 |
|  | D8 | 3T Siemens TrioTim | 1.0 × 1.0 × 1.2 | 3.0/ 2300/ 900 | 0.9 × 0.9 × 5.0 | 90/9000/ 2500 | 127 |
|  | D9 | 3T Siemens TrioTim | 1.1 × 1.1 × 1.2 | 3.0/ 2300/ 900 | 0.9 × 0.9 × 5.0 | 90/9000/ 2500 | 10 |
|  | D10 | 3T Siemens Skyra | 1.0 × 1.0 × 1.2 | 3.0/ 2300/ 900 | 0.9 × 0.9 × 5.0 | 91/9000/ 2500 | 29 |
|  | D11 | 3T Siemens Skyra | 1.1 × 1.0 × 1.2 | 3.0/ 2300/ 900 | 0.9 × 0.9 × 5.0 | 91/9000/ 2500 | 13 |
|  | D12 | 3T Siemens Verio | 1.0 × 1.0 × 1.2 | 3.0/ 2300/ 900 | 0.9 × 0.9 × 5.0 | 91/9000/ 2500 | 32 |
|  | D13 | 3T GE Signa HDxt | 1.0 × 1.0 × 1.2 | 3.0/ 7.0/ 400 | 0.9 × 0.9 × 5.0 | 154/11002/ 2250 | 2 |

Table B.1 lists the acquisition parameters for each dataset. All voxel sizes are given to the nearest 0.1mm while all TE/TR/TI times are given to the nearest 0.1ms. In the dataset column, Chal. refers to the OOD (Challenge) dataset. The total column refers to the number of images included across all experiments. In the Institute column, for the CVD and MSS3 datasets, CCBS refers to Centre for Clinical Brain Sciences at The University of Edinburgh. For the Challenge dataset, UTC refers to University Medical Center Utrecht, SING refers to NUH Singapore and AMS refers to VU Amsterdam. For the ADNI dataset, the institute codes Dx(x) refer to the following ADNI centres: [**D1**: 002; **D2**: 053; **D3**: 006; **D4**: 010, 018, 019, 031, 100; **D5**: 012; **D6**: 130; **D7**: 013; **D8**: 011, 014, 022, 023, 024, 032, 035, 041, 073, 082, 116, 135, 941; **D9**: 067; **D10**: 033, 036, 137; **D11**: 037; **D12**: 009, 068, 072, 114, 123, 141, 153; **D13**: 016] Further information regarding the ADNI dataset can be found at https://adni.loni.usc.edu/.

## Appendix C. Patch-wise scores and WMH UQ coverage metrics for the nnUNet variants

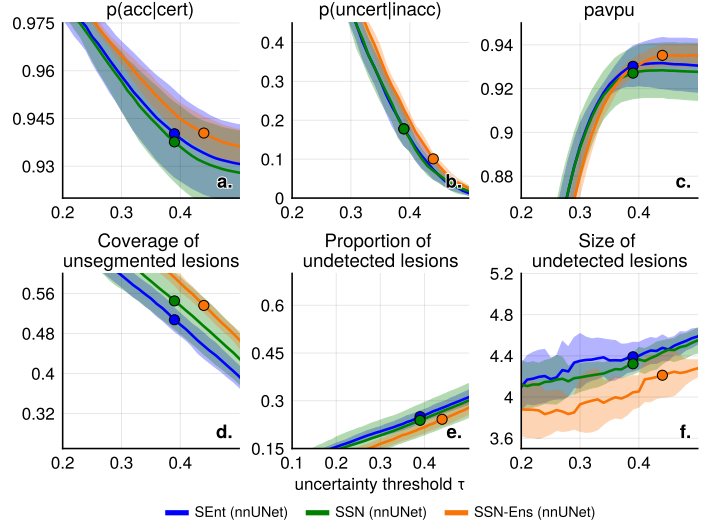

Figure (C.1) (a-c) Patchwise uncertainty metrics for the nnUNet variants of the SEnt, SSN and SSN-Ens models as $\tau$ changes for the OOD (Challenge) dataset. All metrics computed with a sliding window size of $4^3$ voxels and accuracy threshold of 0.8. (a) Probability that a patch is accurate given that it is certain. (b) Probability that a patch is uncertain given that it is inaccurate. (c) PAvPU metric. (d-f) Lesion Instance Coverage metrics for the OOD (Challenge) dataset. Instances are 3D connected components in the ground truth WMH mask. (d) Coverage of unsegmented lesions: The mean proportion of unsegmented instances which are deemed uncertain. (e) Proportion of undetected lesions: The proportion of unsegmented instances for which less than 50% of the instance (or less than 5 voxels, whichever is lower) are deemed uncertain. Dashed line shows the number of instances which are undetected for baseline model SEnt in the WMH segmentation. Most instances are < 10 voxels in volume and undetected. (f) The size of undetected instances.
For all plots, the shaded area denotes the standard deviation over the 6 model runs (equation (12)), and the circles indicate the point at which each method attains an Uncertainty Error Overlap (UEO) of 0.4 on the OOD (Challenge) dataset.

## Appendix D. Uncertainty Map quality metrics per subject

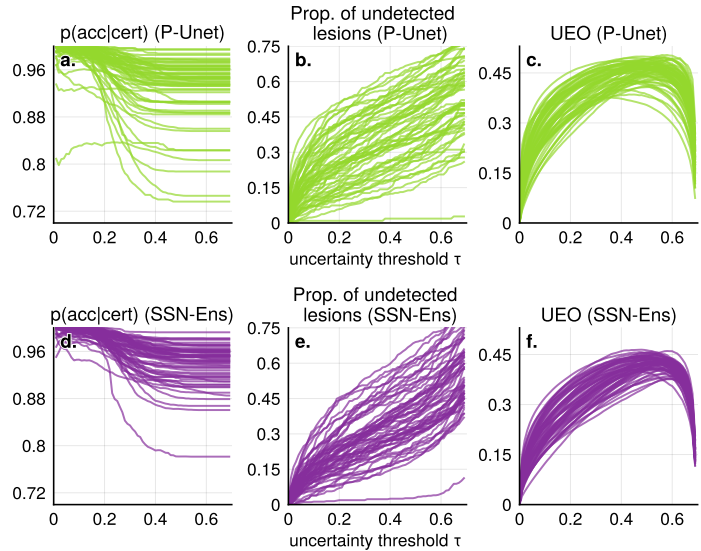

Figure (D.1) Comparison of patchwise metric $p(\text{acc}|\text{cert})$, the proportion of undetected lesions (see figure 5 for reference) and the UEO score per subject for the P-Unet and SSN-Ens methods as the uncertainty threshold $\tau$ changes. One line is plotted per subject. (a), (d) $p(\text{acc}|\text{cert})$ - The probability that a patch of $4^3$ voxels is accurate given that it is certain. (b), (e) The proportion of unsegmented instances for which less than 50% of the instance (or less than 5 voxels, whichever is lower) are deemed uncertain. (c), (e) Uncertainty Error Overlap (UEO).

# Appendix E. Performances of WMH Deep Fazekas score classifiers over all hyperparmeter settings

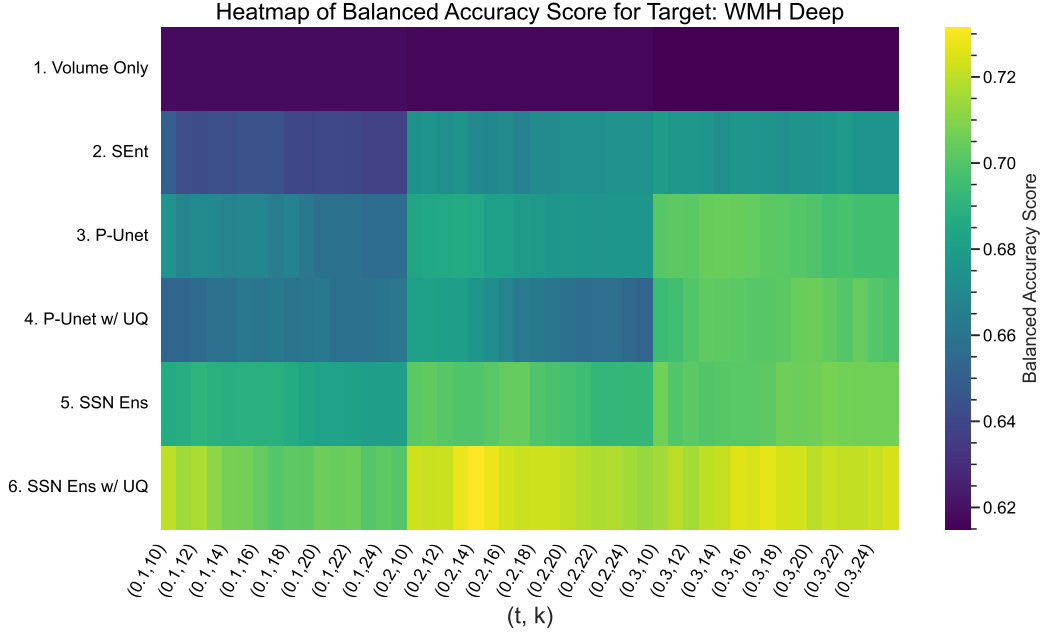

Figure (E.1) Balanced Accuracy Score of the Fazekas classification models for Deep WMH for each setting of hyper-parameters $t$ (the threshold at which model outputs are binarized) and $k$ (the remaining number of features included in the logistic regression model after recursive feature elimination is applied). **Volume Only:** Only the estimated volume from the SEnt model is used as a feature. **SEnt/P-Unet/SSN-Ens:** Spatial and volumetric features are extracted from the segmentation model outputs and used as input features to the Fazekas classifier. **(P-Unet/SSN Ens) w/ UQ** Spatial and volumetric features from the segmentation and uncertainty map are used as input features to the Fazekas classifier. Results are obtained by taking the mean score over 30 randomly sampled train/test splits.

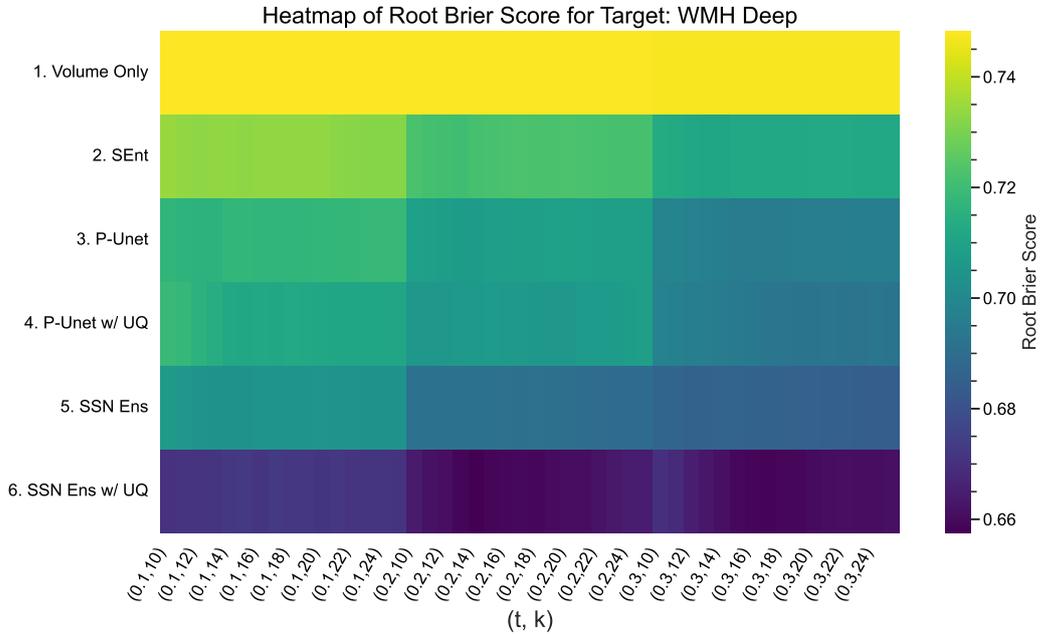

Figure (E.2) Root Brier Score of the Fazekas classification models for Deep WMH for each setting of hyper-parameters $t$ (the threshold at which model outputs are binarized) and $k$ (the remaining number of features included in the logistic regression model after recursive feature elimination is applied). **Volume Only:** Only the estimated volume from the SEnt model is used as a feature. **SEnt/P-Unet/SSN-Ens:** Spatial and volumetric features are extracted from the segmentation model outputs and used as input features to the Fazekas classifier. **(P-Unet/SSN Ens) w/ UQ** Spatial and volumetric features from the segmentation and uncertainty map are used as input features to the Fazekas classifier. Results are obtained by taking the mean score over 30 randomly sampled train/test splits.

## Appendix F. Performances of WMH PV Fazekas score classifiers over all hyperparmeter settings

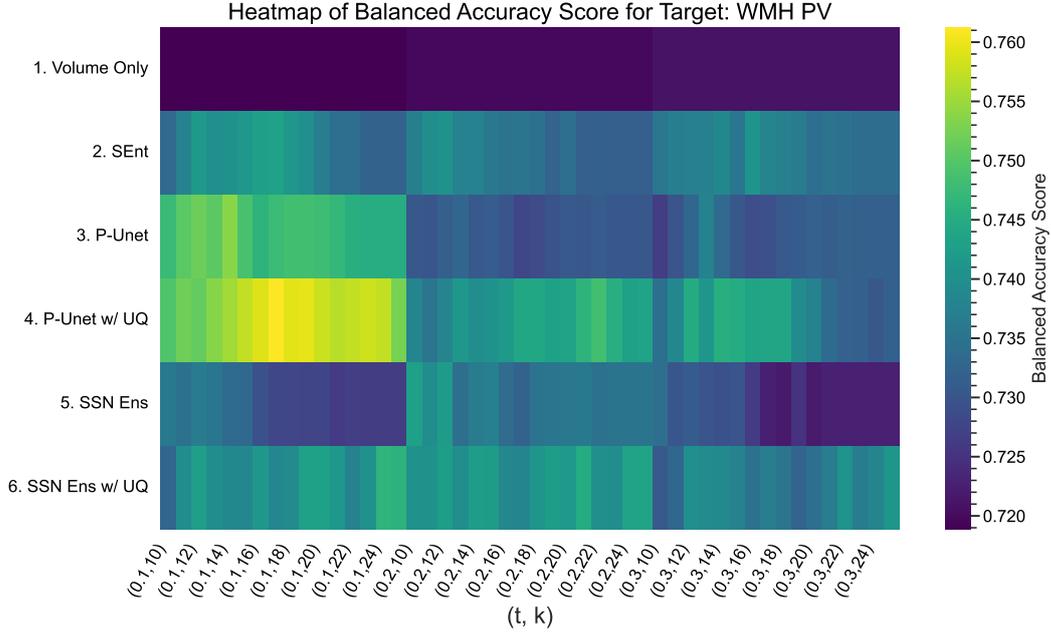

Figure (F.1) Balanced Accuracy Score of the Fazekas classification models for PV WMH for each setting of hyper-parameters $t$ (the threshold at which model outputs are binarized) and $k$ (the remaining number of features included in the logistic regression model after recursive feature elimination is applied). **Volume Only:** Only the estimated volume from the SEnt model is used as a feature. **SEnt/P-Unet/SSN-Ens:** Spatial and volumetric features are extracted from the segmentation model outputs and used as input features to the Fazekas classifier. **(P-Unet/SSN Ens) w/ UQ** Spatial and volumetric features from the segmentation and uncertainty map are used as input features to the Fazekas classifier. Results are obtained by taking the mean score over 30 randomly sampled train/test splits.

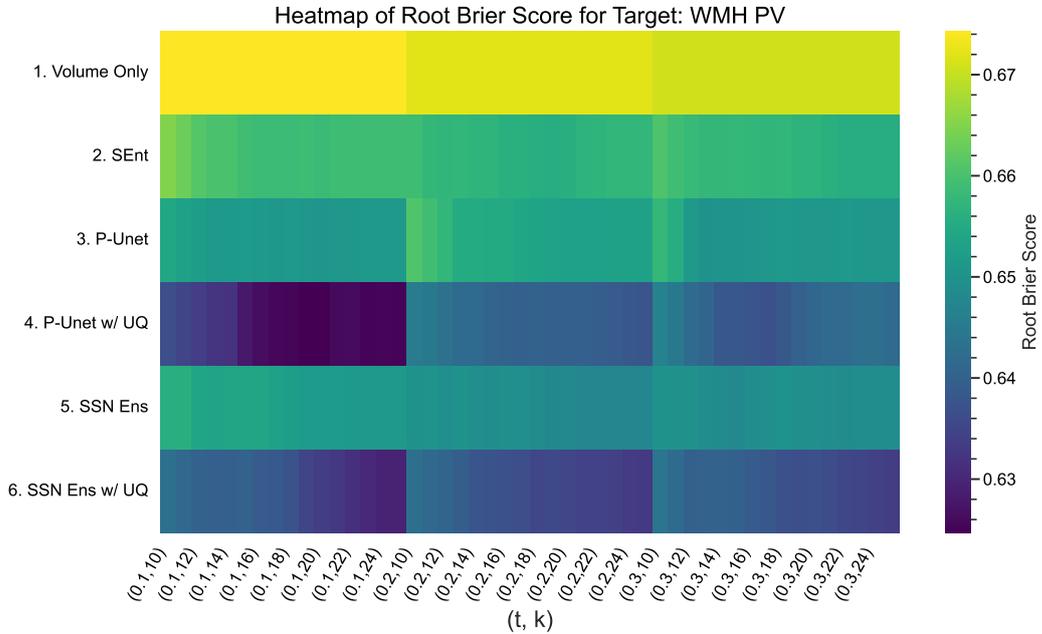

Figure (F.2) Root Brier Score of the Fazekas classification models for PV WMH for each setting of hyper-parameters $t$ (the threshold at which model outputs are binarized) and $k$ (the remaining number of features included in the logistic regression model after recursive feature elimination is applied). **Volume Only:** Only the estimated volume from the SEnt model is used as a feature. **SEnt/P-Unet/SSN-Ens:** Spatial and volumetric features are extracted from the segmentation model outputs and used as input features to the Fazekas classifier. **(P-Unet/SSN Ens) w/ UQ** Spatial and volumetric features from the segmentation and uncertainty map are used as input features to the Fazekas classifier. Results are obtained by taking the mean score over 30 randomly sampled train/test splits.

## Appendix G. Performances of QC classifiers over all hyperparmeter settings

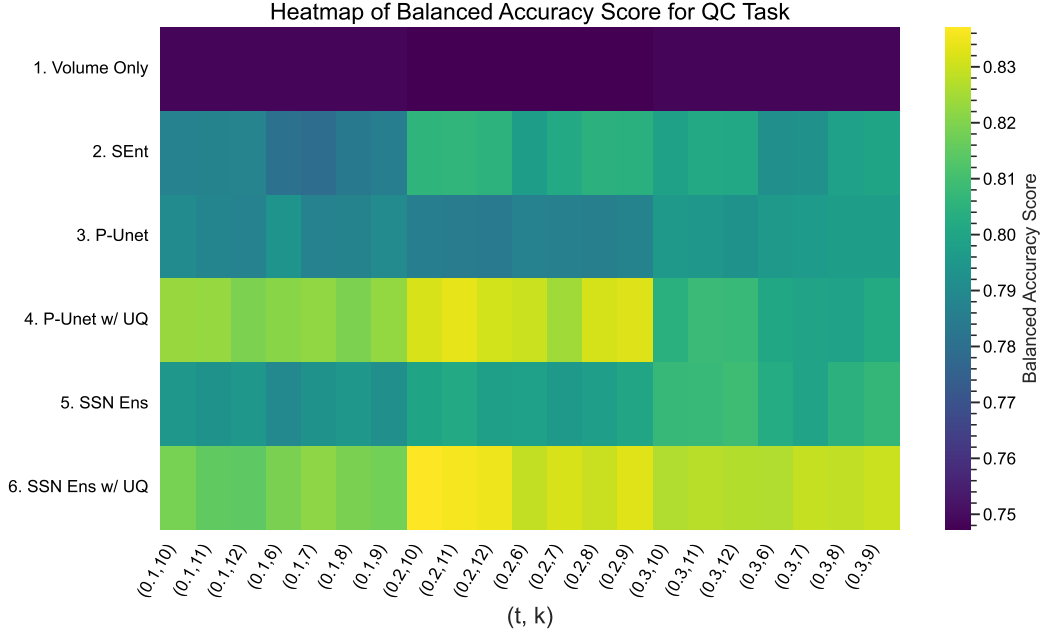

Figure (G.1) Balanced Accuracy Score of the QC classification models for each setting of hyper-parameters $t$ (the threshold at which model outputs are binarized) and $k$ (the remaining number of features included in the logistic regression model after recursive feature elimination is applied). **Volume Only:** Only the estimated volume from the SEnt model is used as a feature. **SEnt/P-Unet/SSN-Ens:** Spatial and volumetric features are extracted from the segmentation model outputs and used as input features to the QC classifier. **(P-Unet/SSN Ens) w/ UQ** Spatial and volumetric features from the segmentation and uncertainty map are used as input features to the QC classifier. Results are obtained by taking the mean score over 30 randomly sampled train/test splits.

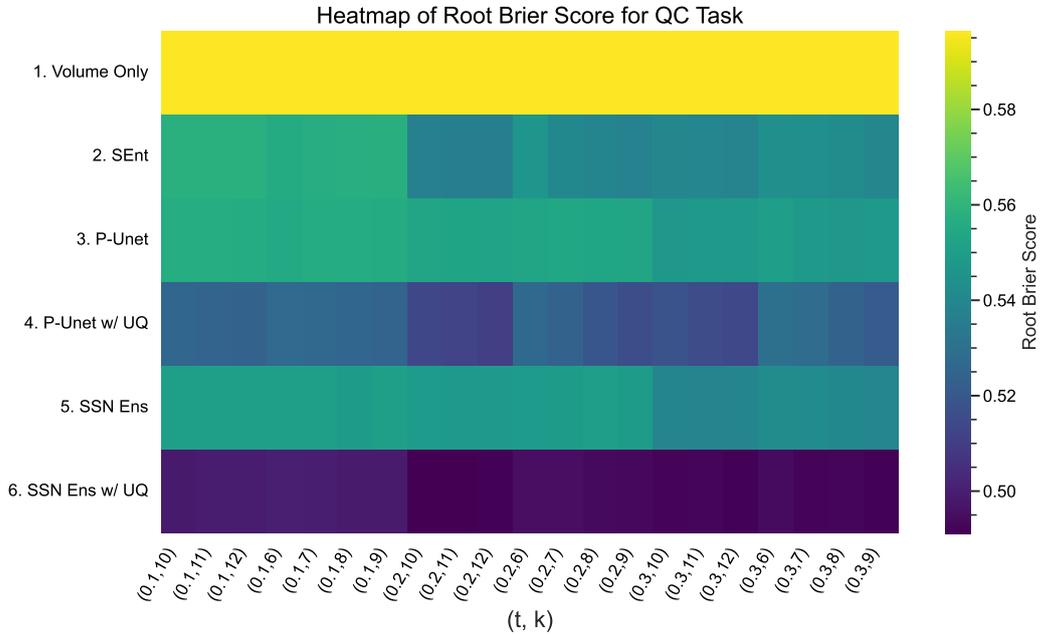

Figure (G.2) Root Brier Score of the QC classification models for each setting of hyper-parameters $t$ (the threshold at which model outputs are binarized) and $k$ (the remaining number of features included in the logistic regression model after recursive feature elimination is applied). **Volume Only:** Only the estimated volume from the SEnt model is used as a feature. **SEnt/P-Unet/SSN-Ens:** Spatial and volumetric features are extracted from the segmentation model outputs and used as input features to the QC classifier. **(P-Unet/SSN Ens) w/ UQ** Spatial and volumetric features from the segmentation and uncertainty map are used as input features to the QC classifier. Results are obtained by taking the mean score over 30 randomly sampled train/test splits.

## Appendix H. Definition of Fazekas categories

The four Fazekas categories for each region may be briefly described as the following:

- 0: Absent or very minimal WMH.

- 1: WMH are discrete and diffuse. For DWMH punctiform, for PVWMH a very thin lining of WMH around the ventricles.

- 2: Substantial areas of WMH in both regions, exibiting early stages of confluence, typically presenting as a halo surrounding the ventricles for PVWMH.

- 3: WMH are large and confluent, with extensive confluence across the DWMH. For PVWMH, large WMH are confluent with WMH in the deep area (making separation between DWMH and PVWMH unclear).

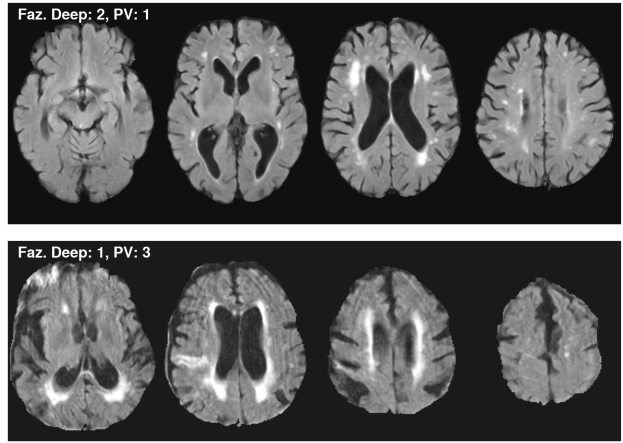

Figure (H.1) Fazekas Categories shown where Deep WMH ≠ PV WMH. Examples taken from the CVD dataset. Top: Deep = 2, PV = 1. Bottom: Deep = 1, PV = 2.

# Appendix I. Further Qualitative Analysis - MSS3 dataset

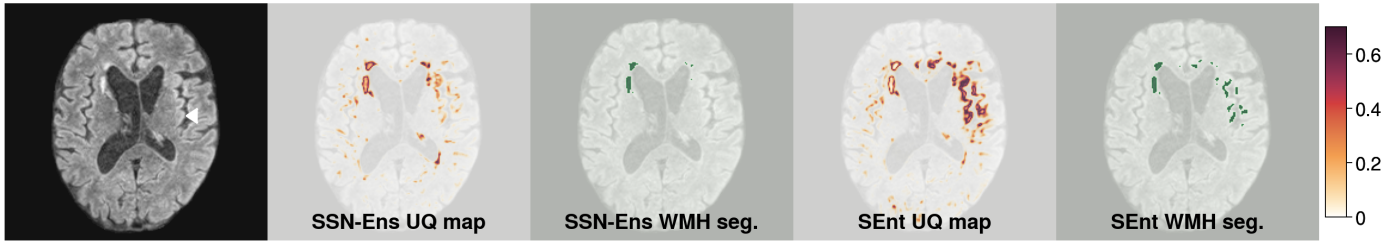

(a) **Incorrect segmentation of the insular cortex on OOD data**. In the right hemisphere the gray matter of the insular cortex is segmented by the SEnt model. White arrow points to a region in the insular cortex segmented as uncertain. The SSN-Ens model is clearly more robust, with less gray matter highlighted as uncertain, and none segmented in this case. The baseline SEnt model does not identify the centre of all segmented regions of the insular cortex as uncertain (see white arrow), demonstrating that the SEnt method does not produce semantically meaningful uncertainty maps on OOD data. Given that this tissue is gray matter, it is not within the aleatoric uncertainty of WMH and reflects the epistemic uncertainty of the model. However, an ideal model would neither segment this area nor deem it uncertain.

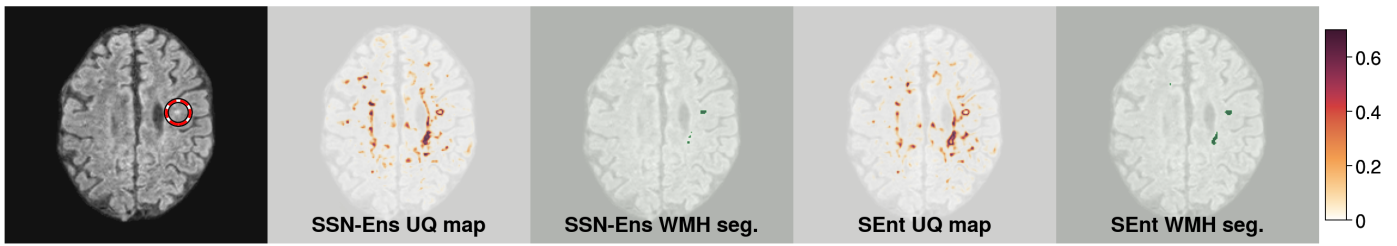

(b) **Small stroke lesions within the spatial distribution of WMH can be misidentified.** Here a small stroke lesion is segmented and not determined as uncertain (except for the boundary) for both models. This stroke lesion is within the distribution of plausible WMH, and if an annotator received no other information than the FLAIR image, may choose to annotate it as WMH (depending on the annotation policy).

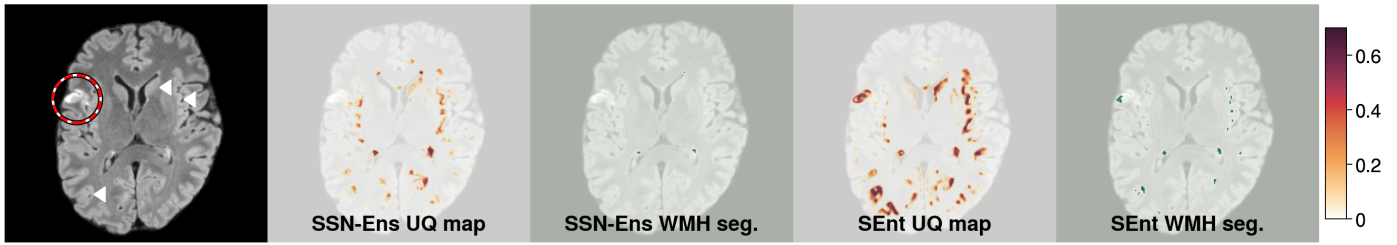

(c) **Difficulty with stroke lesion identification and nuisance gray matter identification for the SEnt model.** The SEnt model highlights part of the stroke lesion as uncertain. However, concerningly, the part of the stroke lesion that is segmented as WMH is not uncertain. The SSN-Ens model does not segment this stroke lesion, nor is it uncertain unlike in other examples. This is most likely due to the stroke lesion appearing outside the typical spatial region of WMH. Both models identify parts of the gray matter, particularly in the insular cortex, as uncertain. The SEnt model further identifies substantial areas of the caudate nucleus (top arrow), insular cortex (middle arrow) and giri (bottom arrow) gray matter regions, as uncertain. This is nuisance information for a human to assess as it is clearly not WMH and limits the utility of uncertainty volumes in downstream tasks. The SSN-Ens model identifies less regions, and with lower intensity, as uncertain, making the uncertainty map easier to threshold (to remove low level uncertainty regions) and leaves less regions to be manually checked by a clinician.

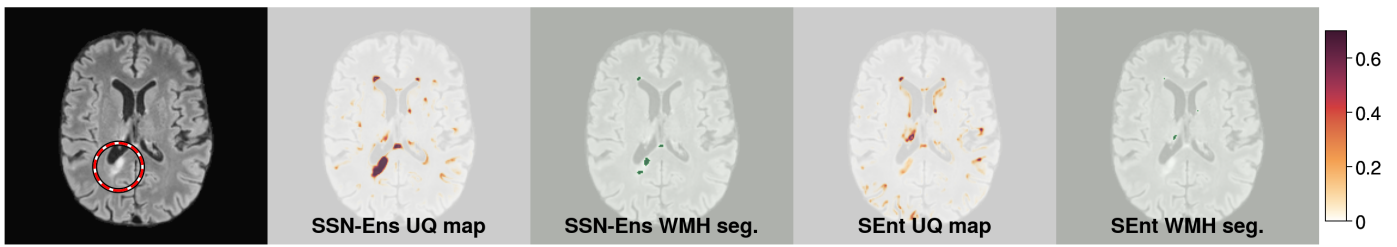

(d) **SSN-Ens identifies a stroke lesion as uncertain. Here a stroke lesion extends from the ventricles.** Both models provide useful outputs here, SEnt (correctly) does not segment this as WMH while SSN-Ens identifies the whole region as highly uncertain, useful for identifying the stroke lesion in the image, albeit at the cost of partially segmenting the region.

Figure (I.1) Qualitative Analysis in the MSS3 dataset (Part 1). Visual assessment of the WMH probability map thresholded at 0.5 (WMH seg.) and UQ maps of the proposed SSN-Ens and baseline SEnt model. Rings denote definite stroke lesions. Colourbars indicate UQ map values.

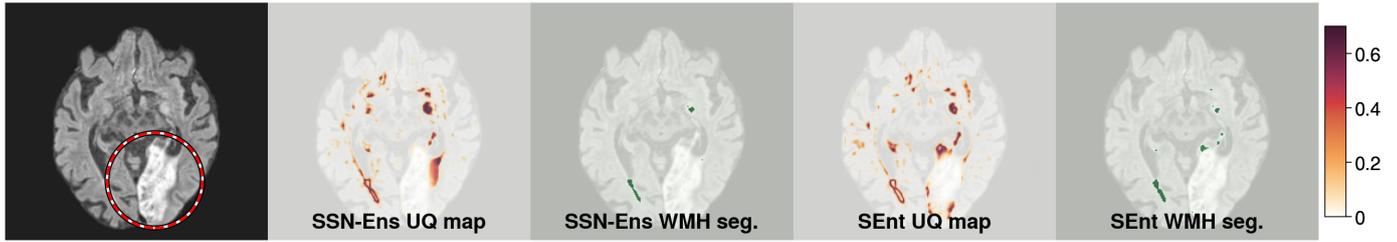

(a) **Separate stroke lesion labels are required for stroke lesions outside the spatial distribution of WMH.** In this figure a large stroke lesion is presented. Due to its size, apparent texture in the image and location, this large hyperintense region is obviously not WMH. Consequently, neither model segments the stroke lesion or identifies it as uncertain. Hence, while the SSN-Ens model in other examples is effective at identifying stroke lesions, these must be at least partially within the spatial distribution of WMH. An effective model for quantifying the uncertainty in both stroke lesions and WMH patologies must include stroke lesion labels during training. In this example SEnt partially (erroneously) segments the lesion, while SSN-Ens provides a valid segmentation.

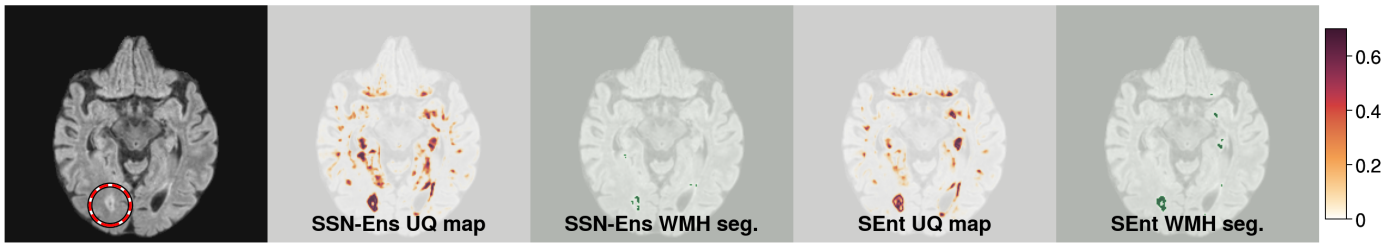

(b) **SSN-Ens identifies a stroke lesion as uncertain, while SEnt does not produce meaningful output.** Both methods segment parts of the stroke lesion as WMH incorrectly, although SSN-Ens only segments a small region, while identifying all of the hyperintense tissue as uncertain. However, the SEnt model only identifies the boundary of the hyperintense region stroke lesion as certain, and is confident that the lesion itself is a WMH. Concerningly, the cavitation within the stroke lesion is uncertain in the SEnt when this could never be a WMH. The SSN-Ens model correctly identifies only the hyperintense area as (highly) uncertain, making the stroke lesion easy to detect from the UQ map.

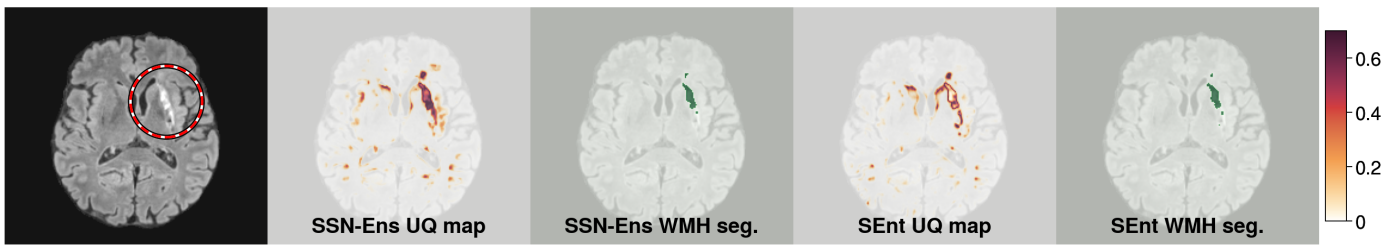

(c) **SEnt fails to identify a stroke lesion (segmented as WMH) as uncertain.** Here both models segment the majority of the stroke lesion as WMH. However, the SSN-Ens model identifies the entire stroke lesion as uncertain while SEnt does not. This allows the SSN-Ens UQ map to be used in a cluster-wise manner, easily identifying regions of the image that may or may not be WMH or stroke lesion. The SEnt UQ map mostly highlights the boundary of the segmented region as uncertain only, which does not tell us whether or not the entire region is possibly a stroke lesion and reduces the utility of the UQ map.

Figure (I.2)  Qualitative Analysis in the MSS3 dataset (Part 2). Visual assessment of the WMH probability map thresholded at 0.5 (WMH seg.) and UQ maps of the proposed SSN-Ens and baseline SEnt model. Rings denote definite stroke lesions. Colourbars indicate UQ map values.

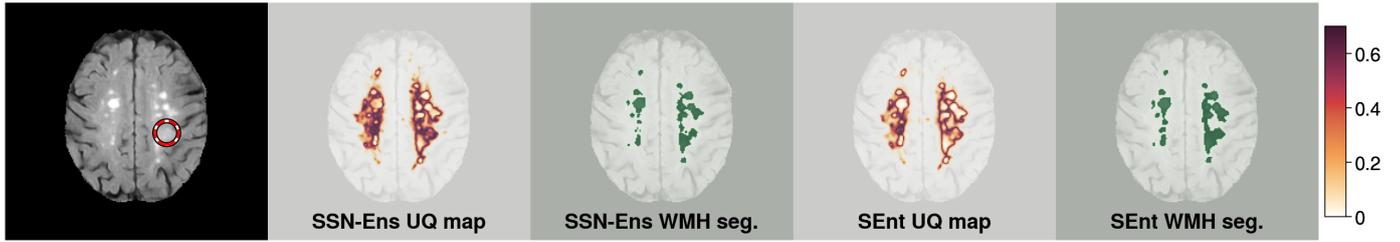

(a) **SSN-Ens uncertain for subtle stroke lesion**. In this image, a subtle stroke lesion, manually detected from a DWI scan, is present in the image. Only SSN-Ens is uncertain about this lesions, while SEnt only highlights the boundary as uncertain.

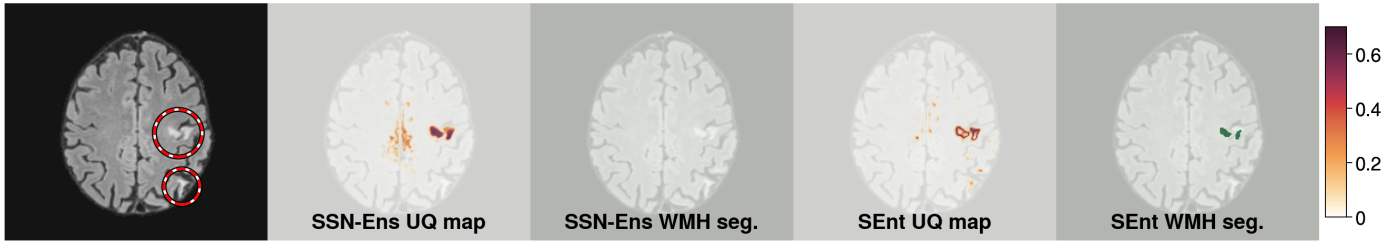

(b) **SEnt incorrectly segments a stroke lesion.** Here SEnt segments a stroke lesion and fails to identify the stroke lesion as uncertain. However, SSN-Ens successfully avoids segmenting the stroke lesion as WMH, while identifying the region as uncertain, making the lesion boundary easily identifiable from the UQ image. However, both models do not detect the stroke lesion at the bottom of the image in the UQ map, likely because this is not within the spatial distribution of WMH.

Figure (I.3) Qualitative Analysis in the MSS3 dataset (Part 3). Visual assessment of the WMH probability map thresholded at 0.5 (WMH seg.) and UQ maps of the proposed SSN-Ens and baseline SEnt model. Rings denote definite stroke lesions. White arrows point to gray matter regions highlighted as uncertain. Colourbars indicate UQ map values.

# Appendix J. Further Qualitative Analysis - MSS3 dataset (using nnUNet variants)

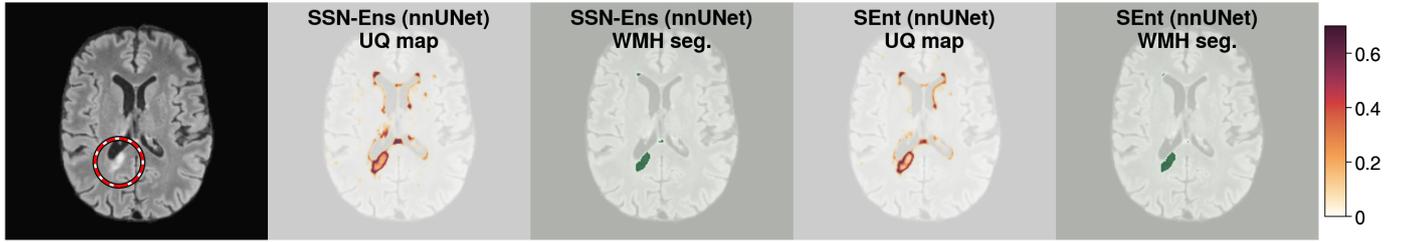

(a) Both models fail to highlight this erroneously segmented stroke lesion as uncertain.

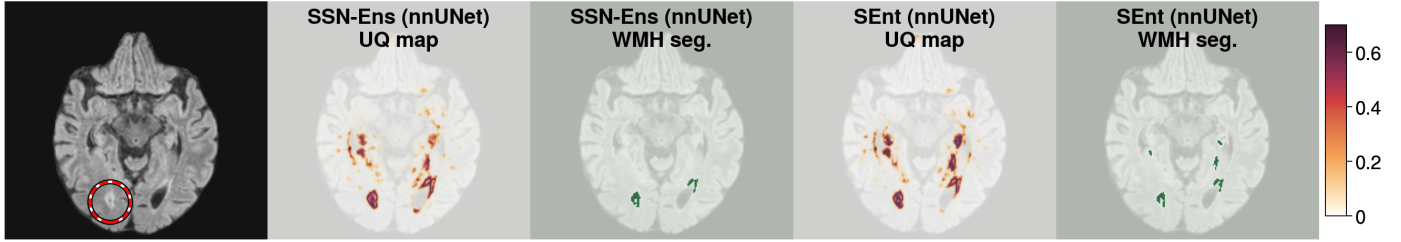

(b) Both models only partially highlight this erroneously segmented stroke lesion as uncertain. SSN-Ens reduces the intensity of uncertainty around gray matter and normal appearing white matter in the image and removes such regions from the segmentation.

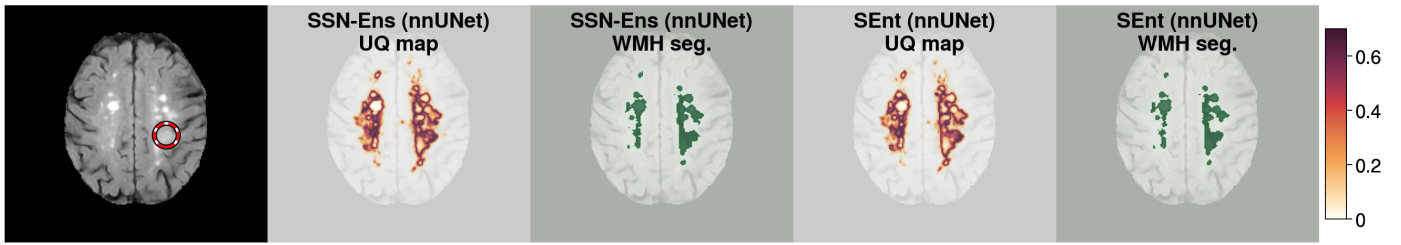

(c) Both models only partially highlight this erroneously segmented stroke lesion as uncertain.

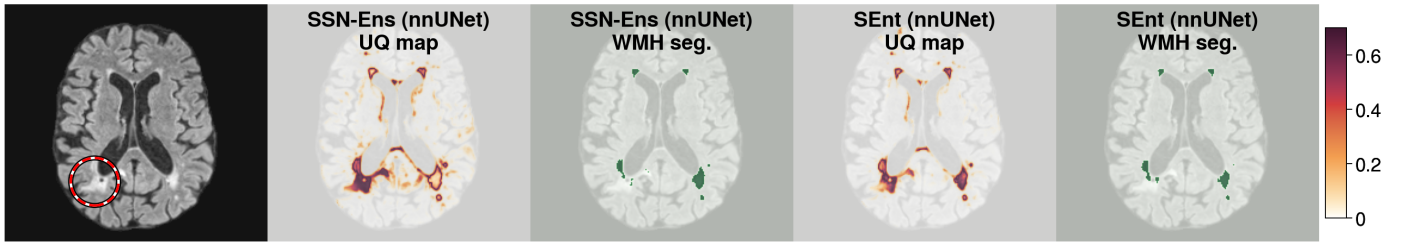

(d) SSN-Ens highlights this stroke lesion has highly uncertain, while the SEnt shows greater confusion between stroke lesions and WMH, with WMH in the right hemisphere uncertain despite accurate segmentation.

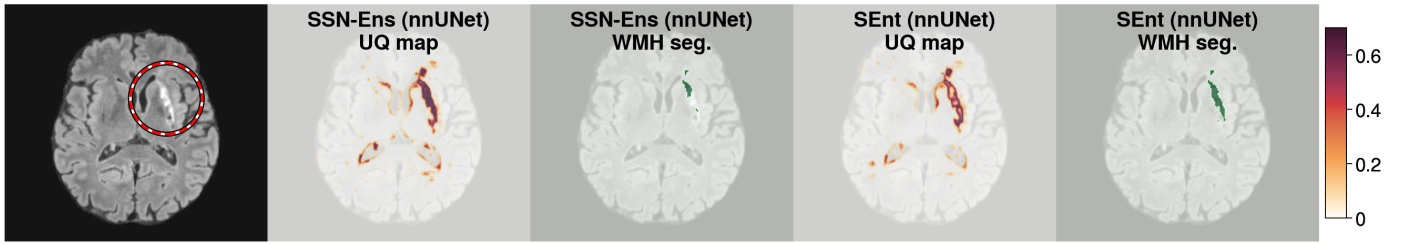

(e) SSN-Ens is uncertain around the stroke lesion which it only partially segments, while SEnt erroneously segments the stroke lesion and is nonetheless certain in the prediction.

Figure (J.1) Qualitative Analysis in the MSS3 dataset. Visual assessment of the WMH probability map thresholded at 0.5 (WMH seg.) and UQ maps of the nnUNet variants of the SSN-Ens and SEnt models. Rings denote definite stroke lesions. Colourbars show uncertainty value in UQ maps.

# Appendix K. Further Qualitative Analysis - Small Lesions in the Challenge Dataset

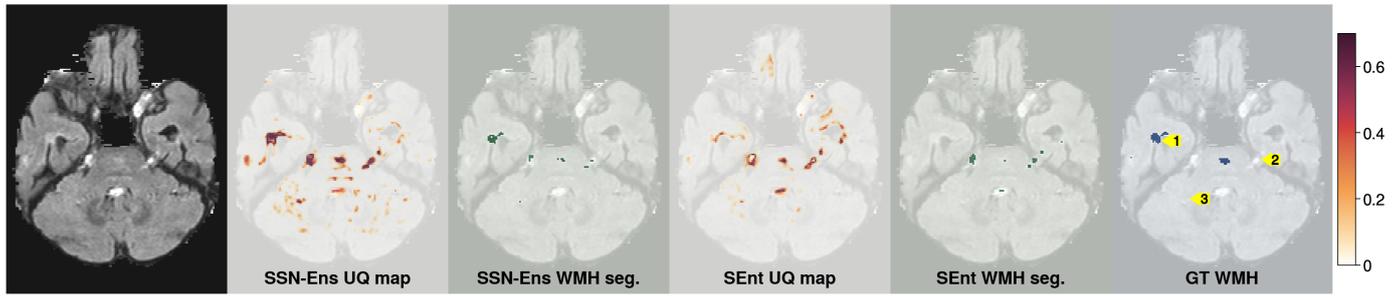

(a) **Arrow 1**: SEnt baseline model fails to segment this WMH, and fails to detect the region as uncertain. **Arrow 2**: Both models introduce false positive WMH here, but SEnt only highlights the boundary of the lesion as uncertain, while SSN-Ens identifies the entire segmented region as uncertain. **Arrow 3**: SSN-Ens model typically introduces regions of the cerebellar hemisphere as uncertain, meaning that low signal in the UQ map should be filtered out prior to visual assessment.

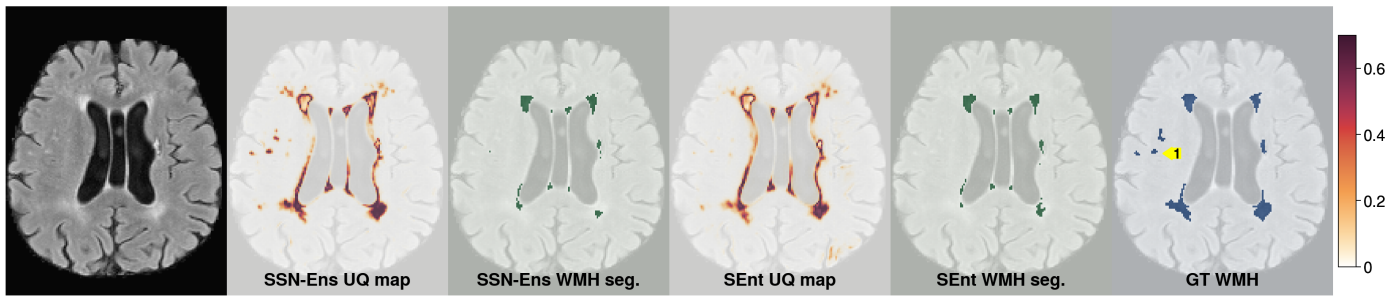

(b) **Arrow 1**: A cluster of 3 WMH of which SEnt fails to segment all 3 while SSN-ENs partially segments a single WMH instance. In the SSN-Ens UQ map all three are detected with high coverage, however SEnt yields very low to zero signal in the UQ map.

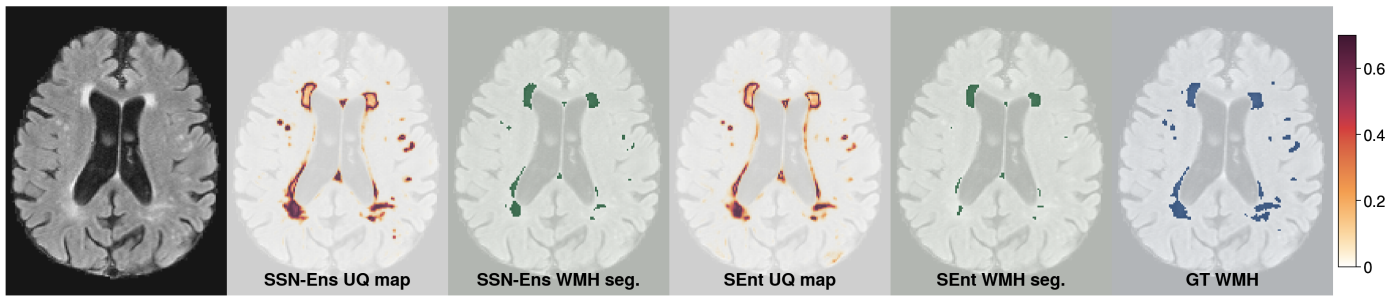

(c) Both methods fail to segment a number of small WMH in this example. However, the UQ map for both methods highlights almost all small WMH. The SSN-Ens yields higher signal in the UQ map for false negatives across the image compared to regions that are not WMH, i.e the uncertainty map is better calibrated.

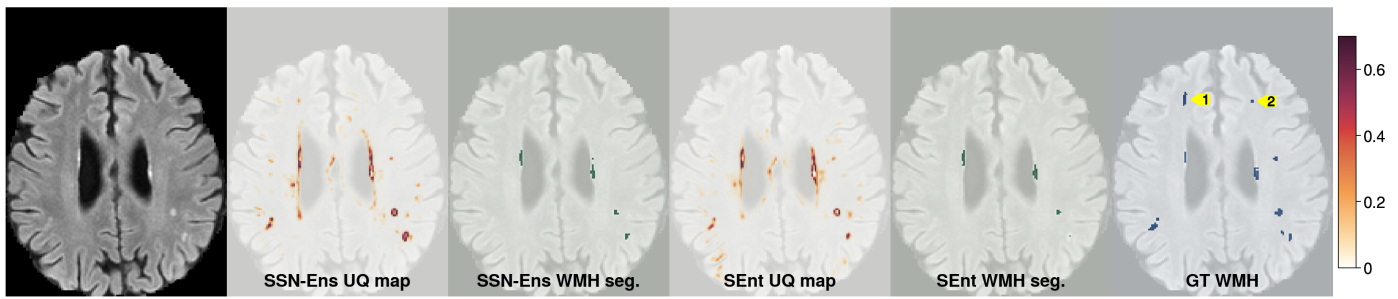

(d) **Arrow 1**: A small WMH is partially detected in the UQ map by SSN-Ens with low signal. SEnt does not identify any part of this WMH as uncertain. **Arrow 2**: Both methods silently fail to either segment this WMH or identify it as uncertain.

Figure (K.1) Qualitative Analysis in the Challenge dataset (part 1). Visual assessment of the WMH probability map thresholded at 0.5 (WMH seg.) and UQ maps of the nnUNet variants of the SSN-Ens and SEnt models. Colourbars show uncertainty value in UQ maps.

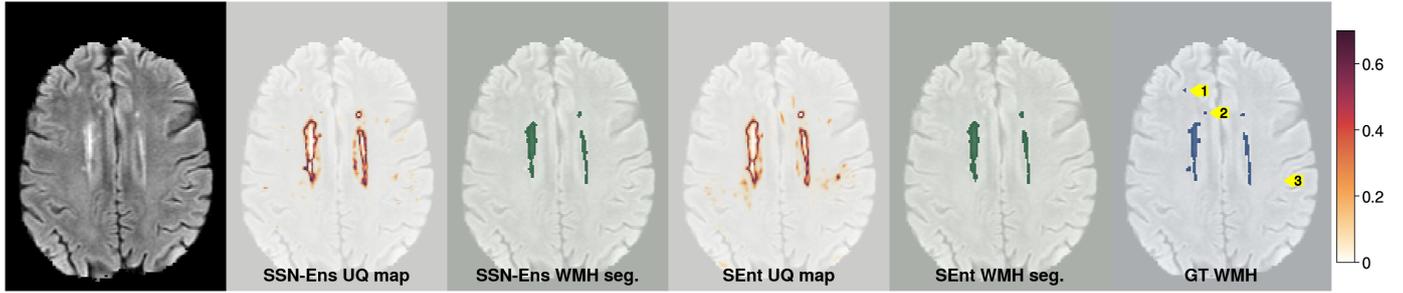

(a) **Arrow 1**: both methods fail to segment this small lesion. SSN-Ens highlights, with very low signal intensity, a few voxels in the centre of the WMH. **Arrow 2**: A repeat of **Arrow 1**. **Arrow 3**: Both methods highlight areas in the uncertainty map that appear to reflect textural changes and gray matter regions in the image, not hyperintensities. While both methods (incorrectly) find a potential WMH at this location, SSN-Ens shows lower signal intensity in the UQ map, demonstrating improved calibration in the uncertainty map.

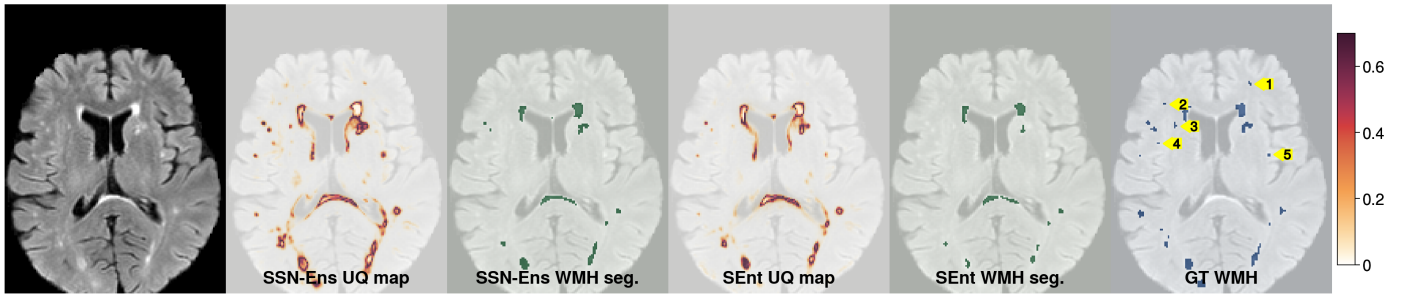

(b) **Arrow 1**: Small WMH is unsegmented by both methods, but detected in the UQ map of SSN-Ens. **Arrow 2**, **Arrow 3**, **Arrow 4**: Repeat of **Arrow 1**. **Arrow 5**: A small WMH, unsegmented by both models. SSN-Ens finds the entire WMH highly uncertain, while SEnt find the WMH partially uncertain, with lower signal intensity in the WMH map.

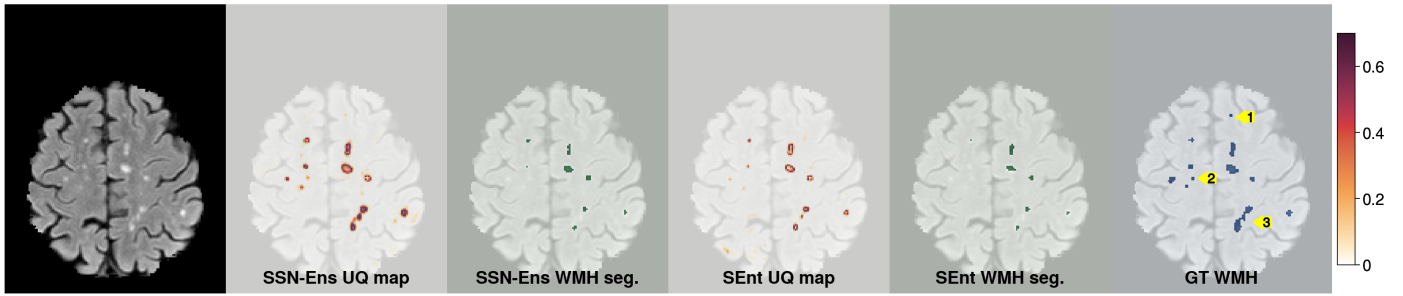

(c) **Arrow 1**: both methods fail to segment this small lesion. SSN-Ens highlights, with very low signal intensity, a few voxels in the centre of the WMH. **Arrow 2**: Small WMH is unsegmented by both methods, but detected in the UQ map of SSN-Ens. **Arrow 3**: Both methods partially segment this cluster of WMH, however SEnt fails to identify as uncertain the regions of the cluster that remain unsegmented.

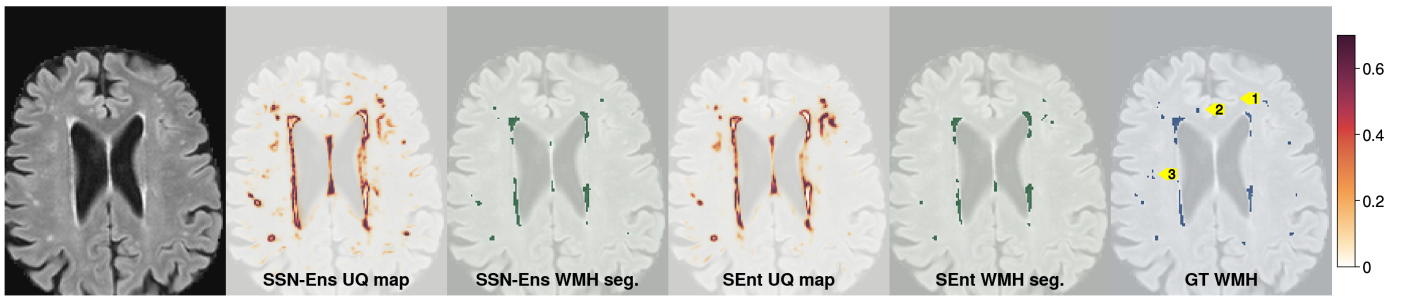

(d) **Arrow 1**: SSN-Ens introduces a number of areas of gray matter as uncertain in the image. This makes segmenting the uncertainty map less useful, as a higher threshold must be set to filter out nuisance regions from the image. This is an effect of domain shift due to MRI aquisition parameter change. In the Challenge dataset, only images collected from the AMS institute (see B.1) with the 3T Philips Ingenuity Scanner exhibit this effect. **Arrow 2**: Both methods fail to identify this WMH in either the segmentation or UQ map. **Arrow 3**: A cluster of 3 small WMH are identified as uncertainty in the SSN-Ens UQ map only. Similarly in the right hand hemisphere, there are three small WMH; only SSN-Ens clearly identifies the top and bottom WMH in the uncertainty map, while neither clearly detect the middle WMH.

Figure (K.2) Qualitative Analysis in the Challenge dataset (part 2). Visual assessment of the WMH probability map thresholded at 0.5 (WMH seg.) and UQ maps of the nnUNet variants of the SSN-Ens and SEnt models. Colourbars show uncertainty value in UQ maps.

# Appendix L. Further Qualitative Analysis - Challenge dataset (using nnUNet variants)

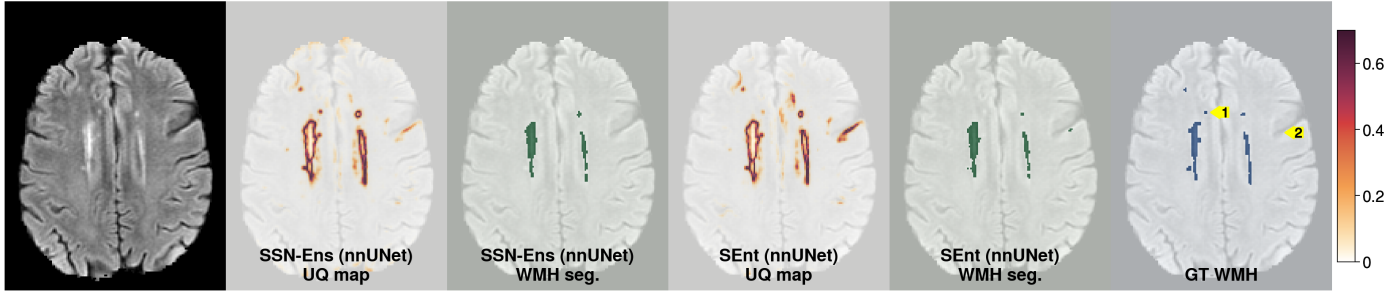

(a) **Arrow 1:** Both methods fail to segment this WMH, however SSN-Ens highlights this region clearly in the uncertainty map, while the SEnt variant is only slightly uncertain in this area. **Arrow 2:** The SEnt variant introduces a number of areas of bright gray matter or normal appearing white matter as uncertain, including this region at the cortex which has also been erroneously segmented. In general the SSN-Ens model greatly reduces the superfluous uncertainty attributed to such regions.

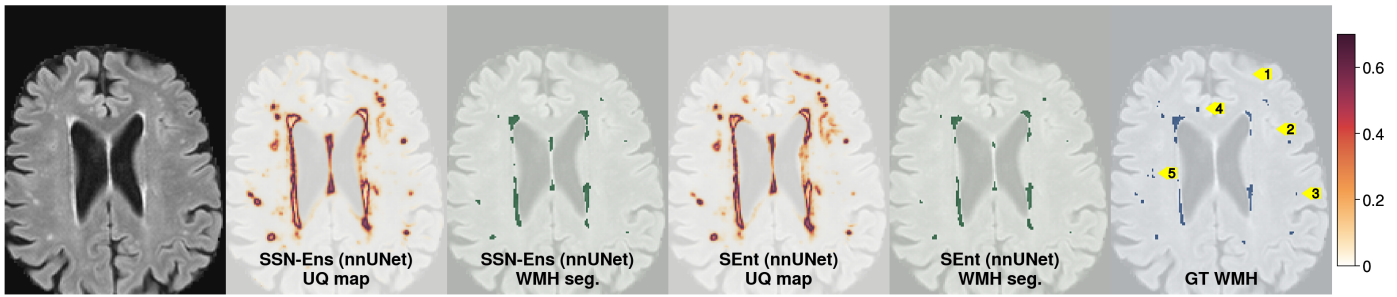

(b) **Arrow 1 and 2:** Similarly to the above example, the SEnt variant introduces a number of bright gray matter or normal appearing white matter regions as uncertain superfluously. In these regions the SSN-Ens variant greatly reduces the uncertainty in these areas. **Arrow 3:** Both methods fail to segment this WMH. SSN-Ens is more uncertain than SEnt, crucially, SSN-Ens is more uncertain around this WMH than it is around arrows 1 and 2, while this is not the case for the SEnt model. **Arrow 4:** An example demonstrating that both methods can silently fail to detect WMH. **Arrow 5:** Both methods fail to segment this cluster of WMH. The SSN-Ens model shows very low uncertainty around this cluster, while the SEnt model does not highlight the region in the uncertainty map. Across the image, SSN-Ens highlights missed WMH or regions in the normal appearing white matter with elevated intensities in the FLAIR as more uncertain than the SEnt variant.

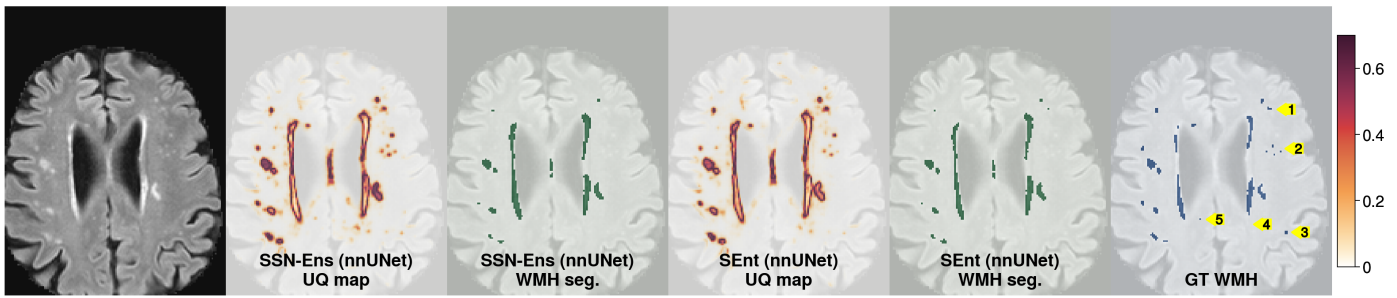

(c) **Arrow 1:** Here is an example of SEnt identifying a WMH in the segmentation that is missed in the segmentation of SSN-Ens; nonetheless, SSN-Ens does identify the WMH in the uncertainty map. **Arrow 2:** Both methods fail to segment this WMH cluster, while SSN-Ens highlights all of the cluster, and with greater intensity than SEnt, in the uncertainty map. **Arrow 3:** Both methods fail to segment this WMH. The SSN-Ens model shows very low uncertainty around this cluster, while the SEnt model does not highlight the region in the uncertainty map. **Arrow 4:** The SSN-Ens model consistently highlights areas within the normal appearing white matter with elevated intensity in FLAIR as uncertain. In this example, both methods highlight this region, with SSN-Ens highlighting the region more strongly. While most likely not WMH in this case, these regions are likely to be areas that may progress to clear WMH in future and hence regions of this type are of high interest for further investigation in longitudinal study. **Arrow 5:** Both methods fail to segment this WMH. The SSN-Ens model shows very low uncertainty around this cluster, while the SEnt model does not highlight the region in the uncertainty map.

Figure (L.1) Qualitative Analysis in the Challenge dataset using nnUnet variants of the SSN-Ens and SEnt methods. Visual assessment of the WMH probability map thresholded at 0.5 (WMH seg.) and UQ maps. Colourbars show uncertainty value in UQ maps.

# Appendix M. Further Qualitative Analysis - model failures in the ADNI dataset (using nnUNet variants)

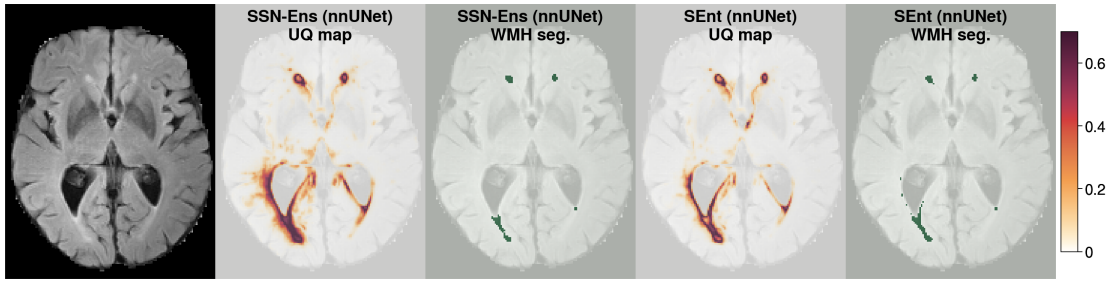

(a) Both models fail to account for the bias field present in this image, in particular over-segmenting the left occipital cap. However, the SSN-Ens model reduces the volume of normal appearing white matter segmented, and where incorrect segmentations are made the model is highly uncertain. However the SEnt variant is confident while incorrectly segmenting normal appearing white matter.

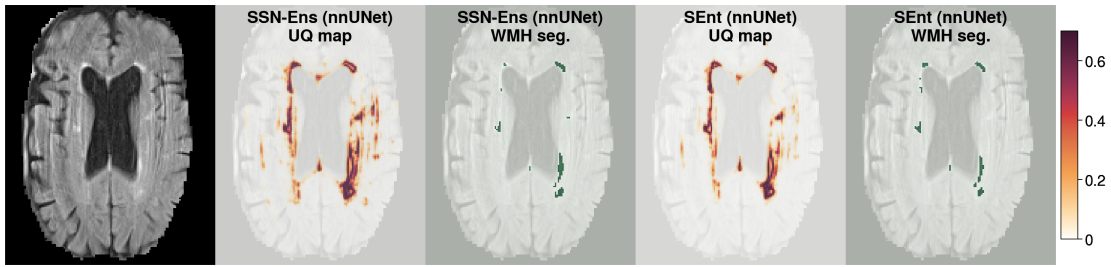

(b) Here the SSN-Ens model is highly uncertain around head motion artifacts in the image, while SEnt shows less uncertainty. Nonetheless, the SSN-Ens model identifies a likely WMH region in the right hemisphere with greater confidence than the SEnt model.

Figure (M.1) Qualitative Analysis in the ADNI dataset using nnUnet variants of the SSN-Ens and SEnt methods. Visual assessment of the WMH probability map thresholded at 0.5 (WMH seg.) and UQ maps. Colourbars show uncertainty value in UQ maps.